\documentclass[prb,amssymb,showpacs,showkeys]{revtex4}
\usepackage{latexsym}
\usepackage{amsfonts,amsbsy,amssymb}

\def\onehalf{\textstyle{\frac{1}{2}}}

\def\gammabol{{\stackrel{\circ}{\Gamma}}{}}
\def\omegabol{{\stackrel{\circ}{\omega}}{}}

\def\Rbol{{\stackrel{\circ}{R}}{}}

\def\abol{{\stackrel{\circ}{a}}{}}

\begin{document}
\title{The Equivalence Principle Revisited}

\author{R. Aldrovandi}

\author{P. B. Barros}

\author{J. G. Pereira}

\affiliation{Instituto de F\'{\i}sica Te\'orica \\
Universidade Estadual Paulista\\ Rua Pamplona 145 \\
01405-900 S\~ao Paulo SP \\ Brazil}

\begin{abstract}
A precise formulation of the strong Equivalence Principle is essential
to the understanding of the relationship between gravitation and
quantum mechanics. The relevant aspects are reviewed in a context
including General Relativity but allowing for the presence of
torsion. For the sake of brevity, a concise statement is proposed
for the Principle: {\em {An ideal observer immersed in a
gravitational field can choose a reference frame in which gravitation
goes unnoticed.}} This statement is given a clear mathematical
meaning through an accurate discussion of its terms. It holds for
ideal observers (time-like smooth non-intersecting curves), but not
for real, spatially extended observers. Analogous results hold for
gauge fields.  The difference between gravitation and the other
fundamental interactions comes from their distinct roles in the
equation of force.
\end{abstract}



\maketitle

\renewcommand{\thesection}{\arabic{section}}
\renewcommand{\thesubsection}{\arabic{subsection}}
 
%%%%%%%%%%%%%%%%%%%%%%
\section{Introduction}
%%%%%%%%%%%%%%%%%%%%%%
 
Despite Synge's forty--years old injunction$^($\cite{Syn60}$^)$ to the
effect that the ``midwife be now buried with appropriate honours'',
the Principle of Equivalence came back to the fore in the last few
years.  Such resurgence is largely due to the difficulties in
reconciling ``local'' General Relativity with ``non--local'' Quantum
Mechanics, $^($\cite{Lam,Man98,MD00,Ahl00}$^)$ but also to possible
violating couplings coming from String Theory.$^($\cite{Dam01}$^)$ A
certain elusiveness in the language, however, seems even today to cast
a shadow on what the Principle really states.  We shall here try to
dispel at least some of that vagueness by proposing a precise
formulation of the Principle which highlights its significance and
limitations.  It should be clear that our concern is the so-called
Strong Principle and not any of its weaker
versions.$^($\cite{Ilioc1}$^)$

In one or another of its forms, the Principle is introduced in every
basic text, and hardly a theme has been the subject of more intense
scrutiny.  In fact, so much has been said about it that we seem to be
reduced to looking for what the Principle {\ \em can\ } state. 
Decades of discussion have brought to light many significant points,
which we intend to bring together into a unified view.  The danger is
worth to be faced that the presentation may seem a pompous listing of
commonplace statements which ``everybody knows''.  This risk is
evident in the version of the Principle we shall use which, put in a
nutshell, reads:
\begin{quote}
An {\em ideal observer} immersed in a gravitational field can choose a
{\em reference frame} in which gravitation goes {\em unnoticed}.
\end{quote}
The apparent simplicity has one merit: it reduces the whole question
to giving clear meanings to the three expressions in italics. 

The key notion is that of ideal observer,$^($\cite{observer}$^)$ which
is a time-like curve, a world-line. Such a curve represents locally,
in well-chosen coordinates, a point-like object in 3-space evolving in
the timelike 4-th direction.  To represent an extended object --- in
particular, a {\em real} observer --- a bunch of world-lines is
necessary, one line for each of its points in 3-space.  A real
observer can know whether (s)he is accelerated or not by making
experiments with extended objects like accelerometers and gyroscopes. 
The gravitational field (that is, the Riemann curvature) will be given
away by the deviations of the world-lines.  We shall see why and how
the Principle holds only for ideal --- point-like in 3-space ---
observers.  It will be seen also that reference frames are better
conceived in terms of fields of vector bases.  Such tetrad fields
include coordinate systems as particular (holonomic) cases.  General,
non-holonomic tetrads provide an extension of the Principle to cases
with torsion.  And it is not difficult to guess in which sense can
gravity ``go unnoticed'': of all the actors with a part in the play,
only can be made to vanish from the scene that one which is
non-tensorial, the connection.$^($\cite{vanishing}$^)$ This raises
another question.  What has been called ``principle of equivalence''
in most recent discussions is a mathematical property, the vanishing
of the Levi-Civita (or Christoffel) connection at a point or along a
curve.  Reduced to that aspect, however, the Principle would not be a
distinctive character of gravitation.  The aptitude to vanish at a
point or along a curve by a convenient choice of ``gauge'' is a
general attribute of connections, shared by gauge potentials. The
meaning and specificity of the Principle is, consequently, involved
with the question of why and how gravitation differs from the other
fundamental interactions.

Spacetime is necessarily conceived as a differentiable manifold:
otherwise we could not even take derivatives, or write differential
equations.  A ``geometry'' consists of structures added to that
initial, basic structure.  The Principle is naturally inbuilt in the
standard geometrical setup of General Relativity.  The use of the
geometrical structure underlying General Relativity is surely
essential for the proper treatment of the subject, but its axiomatic
presentation as surely obscures the experimental evidence for the
existence of that same structure.  The Principle reduces to a theorem. 
Studying it from a more physical point of view helps understanding why
and how gravitation is related to geometry in the way it is.  It is
frequently said that universality ``geometrizes'' the gravitational
interaction.  Universality is absent in the other fundamental
interactions, which have nevertheless also a deep geometrical
connotation.  In which way does the geometrodynamics of gravitation
differ from the geometrodynamics of gauge fields ?  Of course,
gravitation engenders forces of inertial type while the other known
fundamental interactions do not, but why are so different the
geometries related to inertial and to non-inertial forces? 
Understanding the Principle, with its relationship to universality, is
crucial to the understanding of these questions.

The geodesic equation has use only for the symmetric part of the
linear connection involved, which coincides with the whole connection
in the Levi--Civita case.  The vanishing of the symmetric part of a
linear connection {\em at a point} can be achieved by a choice of
coordinates.  Notice that this ``pointwise'' aspect comes from Special
Relativity, more precisely from the Locality
Hypothesis:$^($\cite{Mash00}$^)$ an accelerated observer is supposed
to be equivalent, at each point of its trajectory, to an inertial
observer with the same velocity.  An interesting development has been
the proof$^($\cite{Ilioc,Har95}$^)$ that {\em any} linear connection
can be made to vanish at a previously chosen point by a choice of
tetrad field or, as it is sometimes phrased, by a choice of
non--holonomic coordinates (or still, ``normal frames'').  Such normal
coordinates and/or frames were known to exist at a point and along
smooth non-intersecting curves for symmetric linear connections.  Much
stronger results, concerning general derivatives on tensor algebras,
have been found along the nineties by Iliev.$^($\cite{Ilioc5}$^)$ For
the special cases of covariant derivatives, they encompass both linear
connections --- with torsion or not --- on frame bundles and gauge
potentials on general bundles.

A simple derivation of these results is given below (section
\ref{vanishLinearcon}), whose advantage is that it can be immediately
adapted to gauge theories.  We start with a {\it resum\'e} on tetrad
fields, the Levi-Civita connection and the Weitzenb\"ock connection of
a tetrad.  We exhibit then an explicit tetrad field making a
previously given connection equal to zero at a prescribed point.  In
the presence of torsion, that frame is necessarily non-holonomic.  It
is then shown why a tetrad field which is parallel-transported by a
connection along any prescribed curve produces the vanishing of that
connection along the curve.  A generic tetrad is {\it not}
parallel-transported along a geodesic of the Levi-Civita connection. 
For an ideal observer, however, there will be a tetrad which is.  At
each point $P$ this inertial, free falling frame differs from any
other by a Lorentz transformation which depends on $P$.  These Lorentz
transformations satisfy an interesting equation: the rows of the
Lorentz matrix, seen as vectors, are parallel-transported.

The procedure is then applied to gauge theories (section
\ref{vanishGaugecon}), abelian and non-abelian.  A gauge can be chosen
in which the corresponding connection, the gauge potential, vanishes. 
The difference is obvious: in gravitation a linear connection is at
work, which belongs to the very structure of spacetime.  In gauge
fields, the gauge potential is an ``internal'' connection, acting on
the multiplets of the gauge group.  Dynamically, the difference turns
up in the force equation (section \ref{sec:force}): the gauge field
strength appears explicitly in the Lorentz force, while the
gravitational field strength, the curvature, is absent.

We shall use the notations $\{e_{a}, e^{a}\}$ and $\{h_{a}, h^{a}\}$
for generic tetrad fields, and $\{H_{a}, H^{a}\}$ for tetrads leading
to the vanishing of a connection.  The symbols $(\mu \nu \rho ...)$
and $[\mu \nu \rho ...]$ will indicate symmetrization and
antisymmetrization of included indices in general, as in
${\Gamma}{}^{\lambda}{}_{(\mu \nu)}$ = $\onehalf
\left[{\Gamma}{}^{\lambda}{}_{\mu \nu} + {\Gamma}{}^{\lambda}{}_{\nu
\mu}\right]$ and ${\Gamma}{}^{\lambda}{}_{[\mu \nu]}$ = $\onehalf
\left[{\Gamma}{}^{\lambda}{}_{\mu \nu} - {\Gamma}{}^{\lambda}{}_{\nu
\mu}\right]$.  The ``ball'' notation will indicate the Levi-Civita
connection $\gammabol$ and objects related to it.  Letter $u$ will be
used for the parameter of a curve whose tangent field is $U$, so that
$U^\lambda$ = $ \frac{d x^\lambda}{d u}$ and $U$ = $\frac{d}{d u}$ =
$U^\lambda \partial_\lambda$ ; letter $v$ will be used for the
parameter if the tangent field is $V$, and so on.  Also the standard
notation $\nabla_U$ = $\frac{\nabla\ }{\nabla u}$ will be used.

%%%%%%%%%%%%%%%%%%%%%%%%%%%%%%%%%%%%%%%%%
\section{Metrics, frames and connections}
%%%%%%%%%%%%%%%%%%%%%%%%%%%%%%%%%%%%%%%%%

General Relativity is a metric theory: it takes metrics as fundamental
fields, thereby taking a great distance with respect to the other theories
describing fundamental interactions --- which have connections as the
basic fields.  Given a metric as starting notion, there are a class of
tetrad fields and two connections which have a special significance.

%%%%%%%%%%%%%%%%%%%%%%%%%
\subsection{Frame fields}
%%%%%%%%%%%%%%%%%%%%%%%%%

A coordinate system $\{x^{\mu}\}$ determines, on its domain of
definition, a base $\{\frac{\partial\ }{\partial x^{\mu}}\}$ for the
tangent vector fields and a base $\{dx^\mu\}$ for the covector fields
(1-forms).  These bases are dual, in the sense that $dx^\mu
(\frac{\partial\ }{\partial x^{\nu}}) = \delta^\mu_\nu$.  Such
``holonomic'' bases, related to coordinates, are very particular.  Any
set of four linearly independent fields $\{e_{a}\}$ will form another
base, with a dual $\{e^{a}\}$ such that $e^{a}(e_b) = \delta^a_b$. 
These ``tetrads fields'' are the general linear bases on the spacetime
differentiable manifold.  Their set can be made into another smooth
manifold, the bundle of linear frames.  On the domains both are
defined, the members of a base can be written in terms of the other:
$e_a = e_a{}^\mu \partial_\mu$, $e^{a} = e^{a}{}_\mu dx^\mu$ and
conversely.  The transformations taking $\{e_{a}\}$ into any other
tetrad $\{e'_{a}\}$ constitute the linear group $GL(4, {\mathbb R})$
of all real $4 \times 4$ invertible matrices.  These frames, with
their bundle, are constitutive parts of spacetime, automatically
present as soon as it is supposed to be a differentiable
manifold.$^($\cite{KN96}$^)$

We call {\em linear} connections those related to some subgroup of the
linear group $GL(4, {\mathbb R})$.  They are 1--forms with values in
the Lie algebra of that subgroup.  The Levi--Civita connections of
General Relativity are the most important examples of Lorentzian
connections, which have values in the Lie algebra of the Lorentz
group.  Such connections are ``external'', related to groups acting on
spacetime itself or its tangent spaces, in contraposition to the
``internal'' connections, the vector potentials turning up in gauge
theories.  These have values in the Lie algebra of the gauge group,
which acts on ``internal'' spaces.  A connection defines parallelism
through a covariant derivative.  A vector field is
parallel--transported by a linear connection along a curve if its
covariant derivative vanishes all along the curve.  A field $\phi$
with internal degrees of freedom keeps its direction in internal space
if its covariant derivative, defined by a gauge potential, vanishes
while $\phi$ is displaced along a curve.$^($\cite{AP95b}$^)$

Consider the metric $g$ which has components $g_{\mu \nu}$ in some
holonomic base $\{\frac{\partial\ }{\partial x^{\mu}}\}$:
\begin{equation}
g = g_{\mu \nu} dx^{\mu} \otimes dx^{\nu} = g_{\mu \nu} dx^{\mu}
dx^{\nu}.
     \label{eq:Riemetric}
\end{equation}
There is then a tetrad field $\{h_{a}= h_{a}{}^{\mu}\ \frac{\partial\
}{\partial x^{\mu}}\}$ which relates $g$ to the Lorentz metric $\eta$
by
\begin{equation}
    \eta_{a b}   = g_{\mu \nu} h_{a}{}^{\mu} h_{b}{}^{\nu}.
    \label{eq:gtoeta}
\end{equation}
The $h_{a}$'s are, by this expression, particular linear frames which
are (pseudo-)orthogonal by the metric $g$.$^($\cite{SC92,LL89}$^)$
Each basis $\{h_{a}\}$ constitutes a Lie algebra, fixed either by the
vector fields commutation table
\begin{equation}
    [h_{a}, h_{b}] = f^{c}{}_{a b} h_{c},
    \label{eq:comtable}
\end{equation}
or by its dual expression for the covectors $\{h^{a} = h^{a}{}_{\nu}
dx^{\nu}\}$, Cartan's structure equation
\begin{equation}
    d h^{c} = -\ \onehalf f^{c}{}_{a b}\ h^{a} \wedge h^{b} =   \onehalf \ 
(\partial_\mu h^c{}_\nu - \partial_\nu h^c{}_\mu)\ dx^\mu \wedge dx^\nu.
    \label{eq:dualcomtable}
\end{equation}
The $f^{c}{}_{a b}$'s are the structure coefficients (or anholonomy
coefficients),
\begin{equation}
f^c{}_{a b} = h_a{}^{\mu} h_b{}^{\nu} (\partial_\nu h^c{}_{\mu} - 
\partial_\mu h^c{}_{\nu} ) =  h^c{}_{\mu} [h_a(h_b{}^{\mu}) - 
h_b(h_a{}^{\mu})]  = h^c ([h_a, h_b]). \label{fcab}
\end{equation}
The basis will be anholonomic --- unrelated to any coordinate system
--- if some of the structure coefficients are non-van\-ishing,
$f^{c}{}_{a b} \ne 0$ for some $a, b, c$. The frame
$\{\frac{\partial\ }{\partial x^{\mu}}\}$ has already been presented
as holonomic precisely because their members commute with each other. 
If $f^{c}{}_{a b}$ = $0$, then $d h^{c} = 0$ implies the local
existence of functions (coordinates) $y^c$ such that $h^{c}$ = $d
y^c$.  In that case, the $g_{\mu \nu}$'s would be simply the
components of the Lorentz metric $\eta$ in the coordinate system
$\{x^\mu\}$.

The components of basis $\{h_a, h^b\}$ members with respect to the
base $\{\partial_{\mu}, dx^{\nu}\}$ satisfy
\begin{equation}
h^{a}{}_{\mu} h_{a}{}^{\nu} = \delta_{\mu}^{\nu}; \ \ h^{a}{}_{\mu}
h_{b}{}^{\mu} = \delta^{a}_{b};
    \label{eq:tetradprops1}
\end{equation}
\begin{equation}
g_{\mu \nu} = \eta_{a b}\ h^{a}{}_{\mu} h^{b}{}_{\nu}.
    \label{eq:tettomet}
\end{equation}
Indices can be raised and lowered in a consistent way. Thus, for example, 
\begin{equation}
g^{\mu \nu} = \eta^{a b} h_{a}{}^{\mu} h_{b}{}^{\nu} ; \ \ \eta^{a b}
= g^{\mu \nu} h^{a}{}_{\mu} h^{b}{}_{\nu}; \ \ g_{\mu \nu} = h_{a \mu}
h^{a}{}_{\nu} , \ \ {\textnormal{etc}}.
    \label{eq:other}
\end{equation}

A tensor with components $T^{\lambda_1 \lambda_2 \lambda_3 ...}_{\mu_1
\mu_2 \mu_3 ...}$ in the holonomic base will have, seen from a tetrad
frame $\{h_a\}$, components given by contractions with the tetrad
components.  In particular, Eq.(\ref{eq:tettomet}) tells us that the
metric $g$, seen from the tetrad frame, is just the Lorentz metric. 
This does {\it not} mean that the frame is inertial, because the
metric derivatives --- which appear in the forces and accelerations
--- are not tensorial. To define derivatives which are covariant, it
is essential to add connections $\Gamma^\lambda{}_{\mu \nu}$, whose
non-tensorial behavior in the first two indices compensate the
non-tensoriality of the usual derivatives. Connections obey in
consequence a special law, given below [for example, in
Eqs.(\ref{eq:omegagain}), (\ref{eq:gamtomegabol}) and
(\ref{eq:omega})].  Furthermore, Eq.(\ref{eq:tettomet}) holds for
other tetrad fields.  In effect, another set $\{h_{a'}\}$ can be given
such that $g_{\mu \nu} = \eta_{a b}\ h^{a}{}_{\mu} h^{b}{}_{\nu}\ =
\eta_{c' d'}\ h^{c'}{}_{\mu} h^{d'}{}_{\nu}$.  Contracting both sides
of this expression with $h_{e}{}^{\mu}h_{f}{}^{\nu}$, we arrive at
\begin{equation}
\eta_{a b} = \eta_{c' d'}\ 
(h^{c'}{}_{\mu} h_{a}{}^{\mu} ) (h^{d'}{}_{\nu} h_{b}{}^{\nu}).
\end{equation}
This equation says that the matrices with entries 
\begin{equation}
\Lambda^{a'}{}_{b} = h^{a'}{}_{\mu}\ h_{b}{}^{\mu},
\label{Lortetrad}
\end{equation}
which are such that $h^{a'}{}_{\mu} = \Lambda^{a'}{}_{b}\
h^{b}{}_{\mu}$, satisfy
\begin{equation}
 \eta_{c'd'} \ \Lambda^{c'}{}_{a}\ \Lambda^{d'}{}_{b} = \eta_{a b}.
\end{equation}
This is just the condition that a matrix $\Lambda$ must satisfy in
order to belong to the Lorentz group.  Therefore, basis $\{h_{a}\}$ is
far from being unique.  At each point of the Riemannian manifold, it
is determined by $g$ only up to Lorentz transformations in the tetrad
indices $a, b, c, \ldots$ Tetrads provide matrix representations of
the Lorentz group, but with a special characteristic: they are
invertible.  A group element taking some member of the representation
into another can in consequence be written as in (\ref{Lortetrad}), in
terms the initial and final members.  This establishes a deep
difference with respect to the other fundamental interactions,
described by gauge theories.  There are matrix representations in
gauge theories, like the adjoint representation, but their members are
not invertible. We recall that General Relativity can be entirely
written in terms of tetrads.$^($\cite{SC92,LL89}$^)$

%%%%%%%%%%%%%%%%%%%%%%%%
\subsection{Connections}
%%%%%%%%%%%%%%%%%%%%%%%%

Linear connections have a great degree of ``intimacy'' with spacetime
itself precisely because they are defined on the bundle of linear
frames which, as a constitutive part of spacetime, has some specific
properties, not shared by the bundles related to gauge theories.  In
particular, it exhibits soldering, which leads to torsion.  Linear
connections have torsion while gauge potentials have not.  Soldering
comes from the existence of a ``canonical'', or ``solder''
form,$^($\cite{KN96}$^)$ a 1--form taking vectors on the bundle into
the typical tangent fiber, Minkowski space.  Each tetrad field takes
this typical fiber into the spaces tangent to spacetime, providing a
vector base (and a covector dual base) at each point.  Composition of
the solder form and a tetrad frame takes a vector on the bundle into
one of the very members of that frame.  The torsion $T$ of a linear
connection $\Gamma$, seen from a frame, is just the covariant
derivative of the frame. This is to say that a linear connection has
{\em always} torsion.  $T$ is zero for the Levi-Civita connection of a
metric, but its existence has consequences anyhow: the vanishing of
$T$ is at the origin of the well-known cyclic symmetry of the Riemann
tensor components in General Relativity. In a holonomic base the
torsion components are proportional to the antisymmetric part of the
connection components.  Summing up,
\begin{equation}
 T^{\lambda}{}_{\mu \nu} = \Gamma^{\lambda}{}_{\nu \mu} -
 \Gamma^{\lambda}{}_{\mu \nu}\ = -\ 2\ {\Gamma}{}^{\lambda}{}_{[\mu
 \nu]} = h_{a}{}^{\lambda} \ [ \partial_{\mu} \ h^{a}{}_{\nu} -
 \partial_{\mu} \ e^{a}{}_{\nu} + \omega^{a}{}_{b \mu} \ h^{b}{}_{\nu}
 - \omega^{a}{}_{b \nu} \ h^{b}{}_{\mu} ].
    \label{eq:torsion}
\end{equation}
The right-hand side exhibits the mentioned covariant derivative of the
tetrad field, the $\omega^{a}{}_{b \mu}$'s being the connection
components as seen from the tetrad frame. The decomposition into
symmetric and antisymmetric parts is
\begin{equation} %
{\Gamma}{}^{\lambda}{}_{\mu \nu} = {\Gamma}{}^{\lambda}{}_{(\mu \nu)}
+ {\Gamma}{}^{\lambda}{}_{[\mu \nu]} = {\Gamma}{}^{\lambda}{}_{(\mu
\nu)} -\ \onehalf\ {T}{}^{\lambda}{}_{\mu \nu} \label{decomp2}.
\end{equation} %
This fact already tells us that, in the presence of torsion, it will
be impossible to have all the components $\Gamma^{\lambda}{}_{\mu
\nu}$ equal to zero in a holonomic base.

When a metric is present, the condition of metric compatibility is
that the metric be everywhere parallel--transported by the connection,
that is, that the covariant derivative of the metric be zero:
$\nabla_\lambda g_{\mu \nu} = \partial_\lambda g_{\mu \nu} -
\Gamma^\rho{}_{\mu \lambda} g_{\rho \nu} - \Gamma^\rho{}_{ \nu
\lambda} g_{\mu \rho} = 0$, or
\begin{equation}
	\partial_\lambda g_{\mu \nu} = 2\ \Gamma_{(\mu \nu) \lambda}.
\label{compatibility}
\end{equation}
A metric defines a Levi-Civita connection \ $\gammabol{}$, which is
that unique metric-compatible connection which has zero torsion. The
components are the well-known Christoffel symbols
\begin{equation} %
\gammabol{}^{\lambda}{}_{\mu \nu } = {\textstyle
\frac{1}{2}} g^{\lambda \rho} \left[\partial_{\mu} g_{\rho \nu} +
\partial_{\nu} g_{\rho \mu} - \partial_{\rho} g_{\mu \nu} \right].
\label{Christoffel}
\end{equation} %
The symmetry of $\gammabol$ in its last two indices says that $
{\stackrel{\circ}{T}} = 0$.  If $\Gamma$ preserves a metric and is
not its Levi-Civita connection, it will forcibly have $T \neq 0$. 

The difference between two connections is a tensor.  The contorsion
tensor $K$ of the connection $\Gamma$ is defined by
\begin{equation}
K{}^{\lambda}{}_{\mu \nu } = \gammabol{}^{\lambda}{}_{\mu \nu } -
\Gamma^{\lambda}{}_{\mu \nu }. \label{decomp1}
\end{equation}
Metric compatibility (\ref{compatibility}) implies that contorsion is
fixed by the torsion tensor. Inserting into (\ref{decomp1}) the
expression for ${\stackrel{\circ}{\Gamma}}$ obtained from
(\ref{compatibility}) for the three terms in (\ref{Christoffel}), we
obtain
\begin{equation}%
K^{\lambda}{}_{\mu \nu} = {\textstyle \frac{1}{2}}
\left[T^{\lambda}{}_{\mu \nu} + T_{\mu \nu }{}^{\lambda} + T_{
\nu \mu}{}^{\lambda}\right].  \label{contorsion}
\end{equation}%
Decompositions (\ref{decomp2}) and (\ref{decomp1}) are not the same. 
Compared with (\ref{decomp2}), the two last terms in
(\ref{contorsion}) give an extra symmetric contribution of torsion to
$\Gamma$.

The geodesic equation
\begin{equation}
\nabla_U U^{\lambda} \equiv \frac{\nabla\ }{\nabla u}\ U^{\lambda} =
\frac{d}{du}\ U^{\lambda} + \Gamma^{\lambda}{}_{\mu \nu} U^{\mu}
U^{\nu} = 0
    \label{eq:geodesic1}
\end{equation}
defines a self-parallel curve, whose velocity field $U$ is
parallel-transported by $\Gamma$ along the curve itself.  Seen from
the tetrad frame $\{h_{a}\}$, (\ref{eq:geodesic1}) takes the form
\begin{equation}
    \frac{d}{du}\ U^{a} + \omega^{a}{}_{b c} U^{b}
U^{c} = 0\ ,
    \label{eq:geodesic2}
\end{equation}
where 
\begin{equation}
\omega^{a}{}_{b c} = h^{a}{}_{\lambda} \left[h_{c} (h_{b}{}^{\lambda})
+ h^{a}{}_{\lambda}\ {\Gamma}^{\lambda}{}_{\mu \nu}\ h_{b}{}^{\mu}
h_{c}{}^{\nu} \right] = h^{a}{}_{\lambda} \nabla_{c}
(h_{b}{}^{\lambda})
    \label{eq:omegagain}
\end{equation}
is the connection as seen from the tetrad frame. This transformation
law ensures the tensorial behavior of the covariant deri\-vative:
$\nabla_\nu V^\lambda = h^{a}{}_{\nu} h_b{}^\lambda \nabla_a V^b =
h^{\prime a}{}_{\nu} h^\prime_b{}^\lambda \nabla^\prime_a V^{\prime
b}$ and $\nabla_a V^b = \Lambda^c{}_a \Lambda^b{}_d \nabla^\prime_c
V^{\prime d}$.  We shall frequently use (\ref{eq:omegagain}) in the
alternative form
\begin{equation} %
\partial_{\nu} \ h_{b}{}^{\lambda} + \Gamma^{\lambda}{}_{\mu \nu}\
h_{b}{}^{\mu} = h_{a}{}^{\lambda}\ \omega^{a}{}_{b \nu} \ .
    \label{eq:omegaasdevi} 
\end{equation} %
The velocity vector will follow the law valid for tensors: its components
are related by
\begin{equation} %
U^{a} = h^{a}{}_{\lambda}\ U^{\lambda}.  \label{Ua}
\end{equation} %
This means that $U^{a}$ is, in general, an anholonomic velocity,
analogous to the angular velocity of a rigid body: there exists no
coordinate $x^a$ such that $U^{a} = \frac{d x^a}{du}$.  A vector
field, given by components $U^{\lambda}$, can be seen as the
directional derivative $U = U^{\lambda} \partial_\lambda$ = $\frac{d\
}{du}$, just the derivative along its own (local) integral curve
$\gamma$ with parameter $u$.  For time-like curves $u$ can be seen as
the proper time and $U$ can be interpreted as the four-velocity along
$\gamma$.  Each connection defines an acceleration as its the
covariant derivative with respect to proper time,
\begin{equation}
a^\mu = \nabla_U U^{\mu} \equiv \frac{d\ }{du}\ U^\mu +
\Gamma^\mu{}_{\nu \lambda}\ U^\nu U^{\lambda}.  \label{eq:force1}
\end{equation}
For every metric-preserving connection the acceleration is, as in
Special Relativity, orthogonal to the velocity. If $\Gamma$ is the
Levi-Civita connection, (\ref{eq:force1}) is the force equation to
which a test particle submits in General Relativity.

The antisymmetric part of $\omega^{a}{}_{b c}$ in the last two indices can
be computed by using Eqs.~(\ref{eq:torsion}) and (\ref{fcab}).  The result
shows that torsion, seen from the anholonomic frame, includes the anholonomy:
\begin{equation}
T^{a}{}_{b c}  = -\ f^{a}{}_{b c} - (\omega^{a}{}_{b c} -
\omega^{a}{}_{c b}). \label{torsionfrom}
\end{equation}
There is a constraint on the first two indices of $\omega^{a}{}_{b c}$
\ {\em if}\ \ $\Gamma$ preserves the metric. In effect, Eqs. 
(\ref{compatibility}) and (\ref{eq:gtoeta}) lead to
\begin{equation}%
\omega_{a b c} = - \ \omega_{b a c}.
\label{omisLor1}
\end{equation}%
This antisymmetry in the first two indices, after lowering with the
Lorentz metric, says that $\omega$ is a Lorentz connection.  This is
to say that it is of the form
\begin{equation}
\omega = \onehalf\ J_a{}^b \, \omega^a{}_{bc} \, h^c,
\end{equation}
with $J_a{}^b$ the Lorentz generators written in an appropriate
representation.  Therefore, any connection preserving the metric is,
when seen from the tetrad frame, a Lorentz-algebra-valued 1-form.  If
we use Eq.~(\ref{Lortetrad}) and the inverse
$(\Lambda^{-1})^{a}{}_{b'} = h^{a}{}_{\mu} \, h_{b'}{}^{\mu}$ =
$\eta_{b'c'} \, \eta^{ad} \, \Lambda^{c'}{}_{d}$ = $\Lambda_{b'}{}^a$,
we find how the components change under tetrad transformation:
\begin{equation}
\omega^{a'}{}_{b' \nu} = \Lambda^{a'}{}_c\ \omega^{c}{}_{d \nu}
(\Lambda^{-1})^{d}{}_{b'} + \Lambda^{a'}{}_c\ \partial_\nu
(\Lambda^{-1})^{c}{}_{b'}.
\label{omisLor}
\end{equation}
Under change of tetrad, the connection $\omega$ (which is a
metric-preserving $\Gamma$ seen from any tetrad) transforms as a
Lorentz connection.

The Riemannian metric $g = (g_{\mu \nu})$ is a Lorentz invariant.  For
what concerns $g$, any two tetrad fields $\{h_{a}\}$ and $\{h'_{a}\}$
as above are equivalent.  A metric corresponds to an equivalence class
of tetrad fields, the quotient of the set of all tetrads by the
Lorentz group.  The sixteen fields $h^{a}{}_{\mu}$ correspond, from
the field-theoretical point of view, to ten degrees of freedom ---
like the metric --- once the equivalence under the six-parameter
Lorentz group is taken into account.  In simple words, all tetrad
fields related by Lorentz transformations determine the same metric,
which differs from the Lor\-entz metric if and only if the tetrad is
anholonomic.

%%%%%%%%%%%%%%%%%%%%%%%%%%%%%%%%%%%%%%%
\subsection{The Levi-Civita connection}
%%%%%%%%%%%%%%%%%%%%%%%%%%%%%%%%%%%%%%%

The tetrad field $h_{a}$ takes the components (\ref{Christoffel}) of
the Levi-Civita connection,
\begin{equation} %
\gammabol{}^{\lambda}{}_{\mu \nu } = {\textstyle \frac{1}{2}}
\left\{h_b{}^{\lambda}( \partial_{\nu} h^{b}{}_{\mu} + \partial_{\mu}
h^{b}{}_{\nu}) + g^{\lambda \rho}\left[ h_{a \nu} (\partial_{\mu}
h^{a}{}_{\rho} - \partial_{\rho} h^{a}{}_{\mu}) + h_{a \mu}
(\partial_{\nu} h^{a}{}_{\rho} - \partial_{\rho} h^{a}{}_{\nu})
\right] \right\},
\label{gammainh}
\end{equation} %
into those of the so-called spin-connection
\begin{equation} %
    \omegabol^{a}{}_{b \nu} = h^{a}{}_{\lambda}\
    \gammabol^{\lambda}{}_{\mu \nu}\ h_{b}{}^{\mu} +
    h^{a}{}_{\rho} \ \partial_{\nu} \ h_{b}{}^{\rho}, 
    \label{eq:gamtomegabol} 
\end{equation} %
which is simply $\gammabol$ as seen from the frame defined by the
tetrad $\{h_{a}\}$. The expression above can be rewritten as in
(\ref{eq:omegaasdevi}),
\begin{equation} %
\partial_{\nu} \ h_{b}{}^{\lambda} + \gammabol^{\lambda}{}_{\mu \nu}\
h_{b}{}^{\mu} = h_{a}{}^{\lambda}\ \omegabol^{a}{}_{b \nu}.
    \label{eq:omegabolasdevi} 
\end{equation} %
If the tetrad were parallel-transported everywhere, the left-hand side
would vanish.  The metric would in that case reduce to the Lorentz
metric. Indeed, from the ensuing expression
\begin{equation}
 \gammabol^{\lambda}{}_{\mu \nu}\
 =  h_{b}{}^{\lambda}\ \partial_{\nu} h^{b}{}_{\mu}, 
\end{equation} %
it follows that the curvature tensor vanishes.  The spin-connection
consequently measures how much the tetrad field $\{h_{a}\}$ deviates
from parallelism, and how much the metric differs from that of
Lorentz.  We shall see below [see Eq.(\ref{eq:hderiv}) and the
comments around it] that equation (\ref{eq:omegabolasdevi}) actually
encodes the Equivalence Principle.  Combined with (\ref{fcab}),
(\ref{eq:gamtomegabol} ) leads to
\begin{equation} %
\omegabol^{a}{}_{b c} - \omegabol^{a}{}_{c b} = f^{a}{}_{c b}.
\label{omandf}
\end{equation} %
The force equation is now 
\begin{equation} %
	\frac{d}{du}\ U^{a} + \omegabol^{a}{}_{b c} U^b U^c =
	h^{a}{}_{\mu} a^{\mu} = \abol^a.
\label{forcefromframe}
\end{equation} %
The term $\omegabol^{a}{}_{b c} U^b U^c$ is responsible for the
inertial force, seen from the frame itself.  The weak Principle is
implied, of course, by the total absence of the mass in the ``force''
equation.

Equation (\ref{eq:omegabolasdevi}) gives, for any $U$,
\begin{equation}
\nabla_U h_a{}^\lambda = h_c{}^\lambda\ \omegabol^c{}_{a \nu} U^\nu. 
\label{nablah}
\end{equation}
The timelike member $h_{0}$ of a set $\{h_{a}\}$ of vector fields
constituting a tetrad will define, for each set of initial conditions,
an integral curve $\gamma$.  For that curve, $h_{0} = U$,
$h_0{}^\lambda = U^\lambda$ and $U^{a} = h^{a}{}_{\lambda}
h_{0}^{\lambda}$ = $\delta^{a}_{0}$.  The force law will say whether
the frame, as it is carried along that timelike curve, is inertial or
not: (\ref{nablah}) gives, for that curve, the acceleration
\begin{equation} %
\abol^{\lambda}  = \nabla_U\ h_{0}{}^{\lambda} =
h_{c}{}^{\lambda} \ \omega^{c}{}_{0 0}. 
    \label{accel1} 
\end{equation} %
The Fermi-Walker derivative will be
\begin{equation}
\nabla_U^{(FW)} h_a{}^\lambda = \nabla_U h_a{}^\lambda + \abol_a
U^\lambda - U_a \abol^\lambda = h_c{}^\lambda\ \omegabol^c{}_{a \nu}
U^\nu + \abol_a U^\lambda - U_a \abol^\lambda.  \label{FWforh}
\end{equation}
The particular case 
\begin{equation}
\nabla_U^{(FW)} h_0{}^\lambda = \nabla_U h_0{}^\lambda - \abol^\lambda
= 0
\end{equation} %
implies that $h_0 = U$ is kept tangent along the curve.  Equation
(\ref{accel1}) can be obtained, alternatively, by contracting $U^\nu$
with Eq.(\ref{eq:omegaasdevi}) written for $h_{0}$. The expression
for the frame acceleration seen from the frame itself,
\begin{equation}
 \abol^{c} = \omega^{c}{}_{00} \label{ac},
\end{equation}
follows also from Eq.(\ref{forcefromframe}) by inserting $U^{a} =
h^{a}{}_{\lambda} h_{0}^{\lambda}$ = $\delta^{a}_{0}$, but with one
great advantage: the external acceleration appears then clearly as
equal to the inertial acceleration.  We see here in which sense a
frame satisfying Eq.(\ref{eq:gtoeta}) is equivalent to the
gravitational field defined by the metric $g_{\mu \nu}$: it is
accelerated, and its acceleration is just the inertial gravitational
acceleration.

Of course, $\abol^{0} = \omega^{0}{}_{00} = 0$, and only the space
components are nonvanishing.  Seen from another frame $h_{a'} =
\Lambda^{b}{}_{a'} h_b$, the velocity will be $U^{a'} =
\Lambda^{a'}{}_{b} U^b = \Lambda^{a'}{}_{0}$ and
\begin{equation}
 a^{c'} = h^{c'}{}_\rho a^\rho = \Lambda^{a'}{}_{0}
 h_{a'}(\Lambda^{c'}{}_{0}) + \omega^{c'}{}_{a'b'}\Lambda^{{a'}}{}_{0}
 \Lambda^{{b'}}{}_{0} = \Lambda^{c'}{}_{c} a^c.
\end{equation}

%%%%%%%%%%%%%%%%%%%%%%%%%%%%%%%%%%%%%%%%%%%
\subsection{The Weitzen\-b\"ock connection}
%%%%%%%%%%%%%%%%%%%%%%%%%%%%%%%%%%%%%%%%%%%

Each  tetrad field $\{h_{a}\}$ defines a very particular connection, 
the Weitzen\-b\"ock connection whose components are
\begin{equation}
    {\bar \Gamma}^{\lambda}{}_{\mu \nu} = h_{a}{}^{\lambda} \
\partial_{\nu} \ h^{a}{}_{\mu}. 
    \label{eq:weitzen}
\end{equation}
This connection vanishes when seen from frame $\{h_{a}\}$ itself:
\begin{equation} %
{\bar \omega}^{a}{}_{b \nu} = h^{a}{}_{\lambda} \left[ \partial_{\nu}
\ h_{b}{}^{\lambda} + {\bar \Gamma}^{\lambda}{}_{\mu \nu}\
h_{b}{}^{\mu} \right] = h^{a}{}_{\lambda} {\bar \nabla}_{\nu} \
h_{b}{}^{\lambda} = 0.
    \label{eq:gambartombar} 
\end{equation} %
This means that it parallel-transports each vector of the tetrad
$\{h_{a}\}$ everywhere: ${\bar \nabla}_{\nu} \ h_{b}{}^{\lambda} = 0$
(which justifies the name ``teleparallelism'' given to the approach to
gravity based on this connection).  In consequence, it preserves also
the metric $g$: ${\bar \nabla}_{\lambda} \ g_{\mu \nu} = 0$.  It has
vanishing Riemann curvature tensor: ${\bar R}^{\rho}{}_{\sigma \mu
\nu} = 0$ and ${\bar R}^{a}{}_{b c d} = 0$. It has, however, a
non-vanishing torsion ${\bar T}$, actually a mere version of the
structure coefficients of the tetrad frame seen from the holonomic
base:
\begin{equation}
    {\bar T}^{\lambda}{}_{\nu \mu} = h_{a}{}^{\lambda} \left(
    \partial_{\nu} h^{a}{}_{\mu} - \partial_{\mu} h^{a}{}_{\nu}
    \right) = h_{c}{}^{\lambda} f^{c}{}_{a b}\  h^{a}{}_{\mu}
    h^{b}{}_{\nu}.
    \label{eq:barT}
\end{equation}
 As ${\bar \omega}^{a}{}_{b c}
\equiv 0$, a geodesic of the Weitzen\-b\"ock connection has always the
form
\begin{equation}
    \frac{\bar{\nabla}\ }{\nabla u}\ U^{a} = \frac{d}{du}\ U^{a} = 0
    \label{eq:geodesicbar}
\end{equation}
when seen from the tetrad frame.  The geodesic equation
(\ref{eq:geodesic1}) for $\gammabol$, however, acquires in terms of
${\bar \Gamma}$ an aspect of force equation:
\begin{equation}
\frac{d}{du}\ U^{\lambda} + {\bar \Gamma}^{\lambda}{}_{\mu \nu}\
U^{\mu} U^{\nu} = -\ \bar{K}^{\lambda}{}_{\mu \nu}\ U^{\mu} U^{\nu} =
\onehalf \left[{\bar T}_\mu{}^{\lambda}{}_\nu + {\bar
T}_\nu{}^{\lambda}{}_\mu \right] U^{\mu} U^{\nu}.
    \label{eq:geodesic5}
\end{equation}
Using (\ref{decomp1}) for the Weitzen\-b\"ock connection, we identify
its contorsion as
\begin{equation}
{\bar K}^{\lambda}{}_{\mu \nu} = h_a{}^{\lambda}\ \omegabol^a{}_{b
\nu}\ h^b{}_{\mu}.
\label{contorsion4}
\end{equation}
This means that $\omegabol$ is the Weitzenb\"ock contorsion seen from
the tetrad frame $\{h_a\}$: $ \omegabol^a{}_{b c}$ =
${{\stackrel{\circ}{\bar K}}{}}{^a{}_{b c}}$.  And using
(\ref{contorsion}) with (\ref{eq:barT}) in this expression gives
\begin{equation}
 \omegabol^a{}_{b c} = -\ \onehalf \left[f^a{}_{b c} + f_{b c}{}^a + f_{c
 b}{}^a \right], \label{ombol3}
\end{equation}
where the indices are raised and lowered with the Lorentz metric.  The
trivial property $f_{c (ab)} = 0$ implies (\ref{omisLor1}) for
$\omegabol$.  Quite consistently, $\omegabol^a{}_{b c}$ will vanish if
the base $\{h_a\}$ is holonomic.

Summarizing what we have seen in this section: there is a functional
sixfold infinity of tetrad fields related to a given metric as in
Eq.(\ref{eq:tettomet}).  These tetrad fields differ by point-dependent
(that is, local) Lorentz transformations (wherefrom the functional
sixfoldedness).  Holonomic tetrads correspond to the Lorentz metric
itself.  Each tetrad field defines a Cartan-Weitzen\-b\"ock flat
connection.  This connection is a ``vacuum'' of every other
connection.  When alone, the force law reduces to that of Special
Relativity.  Its interest to the Equivalence Principle is consequently
evident.  We shall actually see that, given a connection $\Gamma$, the
free-falling frame along a curve will be a tetrad field whose
Weitzenb\"ock connection coincides with $\Gamma$ along the curve.

%%%%%%%%%%%%%%%%%%%%%%%%%%%%%%%%%%%%%%%%%
\section{The case of a linear connection}
%%%%%%%%%%%%%%%%%%%%%%%%%%%%%%%%%%%%%%%%%
\label{vanishLinearcon}

Consider a general linear connection $\Gamma$ defined on a manifold
$M$.  Choose a point $P \in M$, and around it a coordinate system
$\{x^{\mu}\}$ such that $x^{\mu}(P)$ = $0$.  Such a system will cover
a neighborhood $N$ of $P$ (its coordinate neighborhood), and will
provide dual holonomic bases $\{\frac{\partial\ }{\partial x^{\mu}}, d
x^{\mu}\}$ for vector and covector fields on $N$.  Any other base will
be given by the components of its members in terms of such initial
holonomic bases, as $\{e_{a} = e_{a}{}^{\mu}(x)\frac{\partial\
}{\partial x^{\mu}}, e^{a} = e^{a}{}_{\lambda}(x) dx^{\lambda}\}$.  It
will be enough for our purposes to consider inside $N$ a non-empty
sub--domain $N^{\prime}$, small enough to ensure that only terms up to
first order in the $x^{\mu}$'s can be retained in the calculations.

Let us indicate by $\Gamma^{\lambda}{}_{\mu \nu}(x)$ the components of
$\Gamma$ in the holonomic base, and by $\omega^{a}{}_{b \nu}(x)$ the
components of $\Gamma$ referred to a generic base $\{e_{a} , e^{a}\}$. 
Such components will be related by
\begin{equation} %
\Gamma^{\lambda}{}_{\mu \nu}(x) = e_{a}{}^{\lambda}(x)\
\omega^{a}{}_{b \nu}(x) \ e^{b}{}_{\mu}(x) + e_{c}{}^{\lambda}(x) \
\partial_{\nu} e^{c}{}_{\mu}(x); \label{omtogam}
\end{equation} %
\begin{equation} %
\omega^{a}{}_{b \nu}(x) = e^{a}{}_{\lambda}(x)\
\Gamma^{\lambda}{}_{\mu \nu}(x)\ e_{b}{}^{\mu}(x) + e^{a}{}_{\rho}(x)
\ \partial_{\nu} \ e_{b}{}^{\rho}(x).
    \label{eq:omega}
\end{equation} %
The piece $e_{c}{}^{\lambda}(x) \ \partial_{\nu} \ e^{c}{}_{\mu}(x)$
in (\ref{omtogam}) is the Weitzenb\"ock connection of the tetrad field
$\{e_{a}\}$.  

%%%%%%%%%%%%%%%%%%%%%%%
\subsection{At a point}
%%%%%%%%%%%%%%%%%%%%%%%

Let $\gamma^{\lambda}{}_{\mu \nu}$ = $\Gamma^{\lambda}{}_{\mu \nu}(P)$
be the value of the holonomic components of the connection at the
point $P$.  On the small domain $N^{\prime}$, the connection
components will be approximated by
\begin{equation}
\Gamma^{\lambda}{}_{\mu \nu}(x) = \gamma^{\lambda}{}_{\mu \nu} +
x^{\rho}\ [\partial_{\rho}\Gamma^{\lambda}{}_{\mu \nu}]_{P}
\end{equation}
to first order in the coordinates $x^{\mu}$.  We shall simply exhibit
a particular base in which the connection components vanish at $P$. 
Indeed, choose on $N^{\prime}$ the tetrad field $\{H_{a}, H^{a}\}$
whose components are
\begin{equation}
\ H_{a}{}^{\lambda} = \delta_{a}^{\lambda} - \delta^{\sigma}_{a}
\gamma^{\lambda}{}_{\sigma \rho} x^{\rho} \ ; \ \ H^{a}{}_{\mu} =
\delta^{a}_{\mu} + \delta^{a}_{\lambda} \ \gamma^{\lambda}{}_{\mu \nu}
x^{\nu}.
\label{choice}
\end{equation}
The Cartan structure equation
\begin{equation}
d H^{a} =  -\ {\textstyle {\frac{1}{2}}} \ c^{a}{}_{bc} \ 
H^{b}\wedge H^{c}   
\end{equation}
 for the 1--forms $\{H^{a}$ =
$H^{a}{}_{\mu} \ dx^{\mu}\}$ will, in the present case, be
\begin{equation}
d H^{a} = {\textstyle {\frac{1}{2}}} \ \delta^{a}_{\lambda} [
\gamma^{\lambda}{}_{\mu \nu} - \gamma^{\lambda}{}_{\nu \mu}] \
dx^{\nu} \wedge dx^{\mu}
= \delta^{a}_{\lambda} \gamma^{\lambda}{}_{[\mu \nu]}\
dx^{\nu} \wedge dx^{\mu}.
\end{equation}
We shall eventually replace indices fixed by $\delta^a_{\lambda}$ for
notational convenience. With this convention the structure coefficients
are, to order zero,
\begin{equation}
c^{a}{}_{bc} =  \gamma^{a}{}_{c b} -  \gamma^{a}{}_{bc} = - \ 
2\ \gamma^{a}{}_{[bc]}. \label{c}
\end{equation}
Keeping always expressions only up to first order in the $x^{\mu}$'s, we
find
$$ %
\omega^{a}{}_{b \nu}(x) = H^{a}{}_{\lambda}(x)\
\Gamma^{\lambda}{}_{\mu \nu}(x)\ H_{b}{}^{\mu}(x) + H^{a}{}_{\rho}(x)
\ \partial_{\nu} \ H_{b}{}^{\rho}(x)
$$%
\begin{equation}
= x^{\mu}\ 
[\partial_{\mu}\Gamma^{a}{}_{b \nu}(P)
- \gamma^{a}{}_{c \nu}\ \gamma^{c}{}_{b \mu}]. \label{omegazero}
\end{equation}
We see that at the point $\{x^{\mu} = 0\}$ the connection components
in base $\{H_{a}\}$ vanish: $\omega^{a}{}_{b \nu}(P)$ = $0$.  The
tetrad $\{H_{a}\}$ is such that its Weitzenb\"ock connection
\begin{equation}
{\bar \Gamma}^{\lambda}{}_{\mu \nu} = \gamma^{\lambda}{}_{\mu \nu} -
\gamma^{\lambda}{}_{\sigma \rho} \gamma^{\sigma}{}_{\mu \nu} x^{\rho}
\end{equation} 
coincides with $\Gamma$ at $P$. The curvature and the torsion tensors
at $P$ are
\begin{equation}
R^{\lambda}{}_{\sigma \mu \nu}(P) = 
\partial_{\mu} \Gamma^{\lambda}{}_{\sigma \nu}(P)  - 
\partial_{\nu} \Gamma^{\lambda}{}_{\sigma \mu}(P) + 
\gamma^{\lambda}{}_{\rho \mu} \gamma^{\rho}{}_{\sigma \nu} - 
\gamma^{\lambda}{}_{\rho \nu} \gamma^{\rho}{}_{\sigma \mu},
\end{equation}
\begin{equation}
T^{\lambda}{}_{\mu \nu}(P)  = \gamma^{\lambda}{}_{\nu \mu}  -
\gamma^{\lambda}{}_{\mu \nu}.
\end{equation}

To see what happens to the anholonomy of base $\{H_{a}\}$, we calculate
\begin{equation}
 [H_a, H_b] = \left\{2 \gamma^c{}_{[ab]} + \left[2 \gamma^d{}_{[ab]}
 \gamma^c{}_{d \rho} + \gamma^c{}_{b d} \gamma^d{}_{a \rho} -
 \gamma^c{}_{a d} \gamma^d{}_{b \rho}\right] x^\rho \right\} H_c
\end{equation}
The commutator vanishes at $P$ if the connection is symmetric.  Otherwise, 
the anholonomy is minus the  torsion of  $\Gamma$ at $P$:
\begin{equation}
[H_a, H_b] = 2 \gamma^c{}_{[ab]} H_c.
\end{equation}

Define a Lorentz transformation like that of Eq.(\ref{Lortetrad}):
\begin{equation} %
\Lambda^{a}{}_{b}(x) = H^{a}{}_{\lambda}\ h_{b}{}^{\lambda}. 
\label{goodLor}
\end{equation} %
The result of contracting Eq.(\ref{eq:omegaasdevi}) with
$H^a{}_{\lambda}$ is
\begin{equation} %
H^a{}_{\lambda}\partial_{\nu} \ h_{b}{}^{\lambda} +
\Gamma^{\lambda}{}_{\mu \nu}\ H^a{}_{\lambda} h_{b}{}^{\mu} =
h_{c}{}^{\lambda}\ H^a{}_{\lambda}\ \omega^{c}{}_{b \nu},
\end{equation} %
from which follows
\begin{equation} %
\partial_{\nu} \ \Lambda^{a}{}_{b} - h_{b}{}^{\lambda} (\partial_\nu
H^a{}_{\lambda} - \Gamma^{\mu}{}_{\lambda \nu}\ H^a{}_{\mu}) =
\partial_{\nu} \ \Lambda^{a}{}_{b} - h_{b}{}^{\lambda} \nabla_\nu
H^a{}_{\lambda} = \Lambda^{a}{}_{c}\ \omega^{c}{}_{b \nu}.
\end{equation} %
At the point $P$, $\nabla_\nu H^a{}_{\lambda}$ = $0$,
so that
\begin{equation}
(\Lambda^{-1})^{a}{}_{c} \partial_\nu 
\Lambda^{c}{}_b\ = \omega^{a}{}_{b \nu}.
\end{equation}
The connection $\omega^{a}{}_{b \nu}$, which is $\Gamma$ seen from the
frame $\{h_a\}$, reduces to a gauge vacuum. If another frame
$H^{\prime a}{}_{\mu} = \Lambda^{\prime a}{}_b h^b{}_\mu$ satisfies
the above condition, then also $(\Lambda^{\prime}{}^{-1})^{a}{}_{c}
\partial_\nu \Lambda^{\prime c}{}_b\ = \omega^{a}{}_{b \nu}$. 
Equating both expressions shows that the Lorentz transformation ${\bar
\Lambda} = \Lambda^{\prime} \Lambda^{-1} $ taking $H^{a}{}_{\mu}$ into
$H^{\prime a}{}_{\mu}$ obeys $\partial_\nu {\bar \Lambda}^{c}{}_b =
0$.  Conversely, if $H^{a}{}_{\mu}$ satisfies the above condition,
then so does $H^{\prime a}{}_{\mu}$ = ${\bar \Lambda}^{a}{}_b\
H^b{}_{\mu}$, with ${\bar \Lambda}^{a}{}_b$ constant.  There is
actually a sixfold infinity of tetrads satisfying the above
conditions, which differ from each other by point-independent Lorentz
transformations.

Using the tetrad $H_{a}$, the metric is seen to have, to first order,
components
\begin{equation}
	g_{\mu \nu} = \eta_{a b}\ H^{a}{}_{\mu}\ H^{b}{}_{\nu}
	= \eta_{\mu \nu} + 2 \gamma_{(\mu \nu) \sigma} \ x^{\sigma},
    \label{eq:gmunu}
\end{equation}
with derivatives 
\begin{equation}
\partial_{\lambda} \ g_{\mu \nu} = 2 \gamma_{(\mu \nu)
\lambda} \ .
\end{equation}
This equality, compared with (\ref{compatibility}), tells us that the
metric is, as expected, parallel--transported by $\Gamma$
at $P$.
 
%%%%%%%%%%%%%%%%%%%%%%%%%%%%%%%
\subsubsection{Without torsion}
%%%%%%%%%%%%%%%%%%%%%%%%%%%%%%%

The choice of basis (\ref{choice}) produces the vanishing of the
connection at $P$ even in the presence of torsion.  To make contact
with the standard $T = 0 $ case, let us remark that 
\begin{equation}
d H^{a} = -\ \delta^{a}_{\lambda} \gamma^{\lambda}{}_{[\mu
\nu]} dx^{\mu} \wedge dx^{\nu} = \onehalf\ \delta^{a}_{\lambda}\
T^{\lambda}{}_{\mu \nu} dx^{\mu} \wedge dx^{\nu},
\end{equation}
so that $\{H_{a}\}$ is holonomic at $P$ when $T^{\lambda}{}_{\mu
\nu}(P)$ = $0$.  To recover the usual prescription for the vanishing
of a symmetric connection at a point it is enough to choose the
coordinates
\begin{equation}
 y^{a} = \delta^{a}_{\mu} x^{\mu} + {\textstyle {\frac{1}{2}}} \
 \delta^{a}_{\lambda} \gamma^{\lambda}{}_{(\mu \nu)} \ x^{\nu} x^{\mu}.
\end{equation}
A non--holonomic base is necessary in the presence of torsion, but a
coordinate base suffices in its absence --- which is, of course, the
standard result. Now:
\begin{equation}
 \frac{d}{du}\ y^{a} = \delta^{a}_{\mu}\ U^{\mu} \ +
 \delta^{a}_{\lambda} \gamma^{\lambda}{}_{(\mu \nu)} \ U^{\nu} x^{\mu};
\end{equation}
\begin{equation}
 \frac{d^{2}}{du^{2}}\ y^{a} = \delta^{a}_{\lambda} \left[ \frac{d}{du}
 U^{\lambda} + \gamma^{\lambda}{}_{(\mu \nu)} \ U^{\nu} U^{\mu}
 \right]  + \delta^{a}_{\lambda} \gamma^{\lambda}{}_{(\mu \nu)} \
 x^{\mu} \left[\frac{d}{du} U^{\nu}\right].
\end{equation}
Consequently, at $P$, 
\begin{equation}
\frac{d y^{a}}{du}  = \delta^{a}_{\mu}\ U^{\mu}(P)\ \ ; \ \ 
 \frac{d^{2} y^{a}}{du^{2}}  = \delta^{a}_{\lambda} a^{\lambda}(P).
\end{equation}
Suppose a self--parallel curve goes through $P$ in $N'$ with velocity
$U^{\mu}$.  Then,
\begin{equation}
\frac{d^{2}}{du^{2}}\ y^{a} = -\ \delta^{a}_{\lambda}\ 
\gamma^{\lambda}{}_{(\mu \nu)} \gamma^{\nu}{}_{(\rho \sigma)}
U^{\rho}U^{\sigma} \ x^{\mu}.
\end{equation}
The Lorentz transformation (\ref{goodLor}) reduces in the present case to
\begin{equation}
\Lambda^{a}{}_{b}(x) = H^{a}{}_{\lambda}\ h_{b}{}^{\lambda} = h_{b}
(y^a).
\end{equation}

Seen from the frame $\{H_a\}$, the velocity is
\begin{equation}
U^{a} = (\delta^{a}_{\mu} + \delta^{a}_{\lambda} \
\gamma^{\lambda}{}_{\mu \nu} x^{\nu})\ \frac{dx^{\mu}}{du} =
\frac{d}{du} y^{a} + \onehalf\ \delta^{a}_{\lambda} \
\gamma^{\lambda}{}_{[\mu \nu]} (x^{\nu} U^{\mu} - x^{\mu} U^{\nu}).
\end{equation}
Notice that $U^{a}$ = $H^a(\frac{d}{du})$ is a
non-holonomic velocity if $T \ne 0$. If $T = 0$,
$U^{a}$ = $\frac{d}{du} y^{a}$.

In this standard general-relativistic case, given a geodesic $\gamma$
going through a point $P$ ($\gamma(0) = P$), there is always a very
special system of coordinates (riemannian normal
coordinates,$^($\cite{Syn60}$^)$ or geodesic
coordinates$^($\cite{SC78}$^)$) in a neighborhood $N'$ of $P$ such
that the components of the Levi-Civita connection vanish {\it at the
point $P$}.  The geodesic is a straight line $y^a = c^a u$.  As long
as $\gamma$ traverses $N'$, the {\em ideal} observer will not feel
gravitation: the geodesic equation reduces to the forceless equation
$\frac{d u^a}{du} = \frac{d^2 y^a}{du^2} = 0$. This is an inertial
observer in the absence of external forces. Coupling to source fields
and test particles is given by the minimal coupling prescription:
derivatives go into covariant derivatives. If the components of the
connection vanish, covariant derivatives reduce to usual derivatives. 
All laws of Physics reduce to the expressions they have in Special
Relativity.

Actually, $du = \sqrt{g_{\mu \nu} dx^\mu dx^\nu}$ has been taken as
the curve parameter, and the curve stands on the Riemann space.  For
Special Relativity to hold, it would be necessary to change to the
special relativistic proper time $d\sigma = \sqrt{\eta_{ab} dx^a
dx^b}$ as a parameter.  This change is achieved by a rescaling $dx^a
\rightarrow dx^a \frac{du}{d\sigma}$ in all the coordinates.  This
would correspond to transforming $H^a{}_\mu$ to the tetrad $H^{\prime
a}{}_\mu = \frac{\partial x^a}{\partial x^\mu} \frac{du}{d\sigma}$.

There is actually more, as will be seen in \ref{subsec:Alongacurve}. 
Given any smooth curve, it is possible to find a coordinate system,
defined on a domain $U$, in which the components of the Levi-Civita
connection vanish {\it along the curve}.  And still more: along {\it
any} differentiable curve and {\it any linear connection}, it is
possible to find a local frame, defined on a domain $N'$, in which the
connection components vanish along the curve.  Gravitation {\it seems}
to be absent.  Using that frame an ideal observer, accelerated or not,
can employ Special Relativity.  A real observer will see more.  Only
the connection appears in the geodesic equation, but curvature (and,
eventually, torsion) makes itself present in the Jacobi equation. 
Before going into that, let us say a few words on what happens in the
presence of torsion.

%%%%%%%%%%%%%%%%%%%%%%%%%%%%
\subsubsection{With torsion}
%%%%%%%%%%%%%%%%%%%%%%%%%%%%

Torsion will not add a force, by the last expression in
Eq.(\ref{decomp2}), which leads to
\begin{equation}
 a^{\lambda} = \frac{d}{du} U^{\lambda} + \Gamma^{\lambda}{}_{\mu
 \nu}\ U^{\mu}\ U^{\nu} = \frac{d}{du} U^{\lambda} +
 \Gamma^{\lambda}{}_{(\mu \nu)}\ U^{\mu}\ U^{\nu}.
 \end{equation}
In what concerns the force, only the symmetric part of the connection
contributes and a choice of coordi\-nate system is enough to make the
effect of torsion to vanish.  All the geodesics are consequently fixed
by the symmetric part of the connection, which is by itself a
connection.  Add\-ing a torsion to a symmetric connection does not
change the geodesics.  Nevertheless, these are not the geodesics of\ \
$\gammabol{}^{\lambda}{}_{\mu \nu}$ if $\Gamma{}^{\lambda}{}_{\mu
\nu}$ is metric preserving.  The difference comes from the symmetric
part added by torsion to $\gammabol$, as shown in
Eq.(\ref{contorsion}):
 \begin{equation}
 a^{\lambda} = \frac{d}{du} U^{\lambda} + \Gamma^{\lambda}{}_{(\mu
 \nu)}\ U^{\mu}\ U^{\nu} = \frac{d}{du} U^{\lambda} +
 {\stackrel{\circ}{\Gamma}}{}^{\lambda}{}_{\mu \nu}\ U^{\mu}\ U^{\nu}
 + {\textstyle {\frac{1}{2}}} \ \left[ T_{\mu}{}^{\lambda}{}_{\nu} +
 T_{\nu}{}^{\lambda}{}_{\mu} \right] \ U^{\mu}\ U^{\nu}.
 \end{equation}
There are consequently two distinct coordinate systems:
\begin{description}
\item[(i)] one makes the ``force'' term coming from the Levi--Civita piece
to vanish:

${\stackrel{\circ}{y}}{}^{a} = x^{a} + {\textstyle {\frac{1}{2}}} \
{\stackrel{\circ}{\gamma}}{}^{a}{}_{\mu \nu} x^{\mu} x^{\nu}$;

\item[(ii)] the other causes a metric--preserving ${\bar
{\omega}}{}^{a}{}_{b c}$ to vanish:

${\bar {y}}{}^{a} = x^{a} + {\textstyle {\frac{1}{2}}} \
\left[{\stackrel{\circ}{\gamma}}{}^{a}{}_{\mu \nu} +
T_{\mu}{}^{a}{}_{\nu}(P) + T_{\nu}{}^{a}{}_{\mu}(P) \right] x^{\mu}
x^{\nu}$.
\end{description}
 
%%%%%%%%%%%%%%%%%%%%%%%%%%
\subsection{Along a curve}
%%%%%%%%%%%%%%%%%%%%%%%%%%
\label{subsec:Alongacurve}

The previous results hold for any tetrad whose covariant derivative
vanishes at $P$.$^($\cite{Har95}$^)$   Indeed, expression
(\ref{eq:omegaasdevi}) says that 
\begin{equation}
\partial_{\nu} \ H_{b}{}^{\lambda}(x) + \Gamma^{\lambda}{}_{\mu
\nu}(x)\ H_{b}{}^{\mu}(x) = H_{a}{}^{\lambda}(x)\ \omega^{a}{}_{b
\nu}(x).
\end{equation}  
The condition $\omega^{a}{}_{b \nu}(P) = 0$ is the same as
\begin{equation}
\nabla_{\nu} H_{b}{}^{\lambda}(P) = \partial_{\nu} \
H_{b}{}^{\lambda}(P) + \Gamma^{\lambda}{}_{\mu \nu}(P)\
H_{b}{}^{\mu}(P) = 0.
\label{tovanish3}
\end{equation}
Expressions (\ref{choice}) provide a first-order solution of 
\begin{equation}
 H_{b}{}^{\lambda}(x)  \partial_{\nu} \ H^{b}{}_{\mu}(x)  = 
\Gamma^{\lambda}{}_{\mu \nu}(x) \label{HWeitzen}
\end{equation}
in the neighborhood $N'$ of $P$.  The left-hand side is just the
Weitzenb\"ock connection ${\bar \Gamma}^{(H) \lambda}{}_{\mu \nu}$ of
the tetrad field $\{H_{a}\}$.

The next natural question is: can the same be done along a curve ? 
Take a differentiable curve $\gamma$ which is an integral curve of a
field $U$, with $U^\mu$ = $\frac{dx^\mu}{du}$ =
$\frac{d\gamma^\mu(u)}{du}$.  The condition for the connection to
vanish along $\gamma$, $\omega^{a}{}_{b \nu}(\gamma(u)) = 0$, will be
\begin{equation} 
U^{\nu} \partial_{\nu} \ H_{a}{}^{\lambda}(\gamma(u)) +
\Gamma^{\lambda}{}_{\mu \nu}(\gamma(u))\ U^{\nu}\
H_{a}{}^{\mu}(\gamma(u)) = 0,
\end{equation}
that is, 
\begin{equation} %
\frac{d}{du} \ H_{a}{}^{\lambda}(\gamma(u)) +
\Gamma^{\lambda}{}_{\mu \nu}(\gamma(u))\ U^{\nu}\
H_{a}{}^{\mu}(\gamma(u)) = 0.  \label{tovanish4}
\end{equation}

This is simply the requirement that the tetrad (each of the four
members $H_a$) be parallel-transported along $\gamma$.  Thus, if the
connection vanishes at an end-point of the curve in a frame, it will
vanish when seen from the parallel displacements of that
frame.$^($\cite{Ilioc}$^)$ For the dual base $\{H^a\}$, the above formula
reads
\begin{equation} %
\frac{d}{du} \ H^{a}{}_{\mu}(\gamma(u)) -
\Gamma^{\rho}{}_{\mu \nu}(\gamma(u))\ U^{\nu}\
H^{a}{}_{\rho}(\gamma(u)) = 0.  \label{tovanish5}
\end{equation}
Any differentiable curve $\gamma$ defines a mapping between tangent
spaces by parallel displacement; this means that, given such a curve
and a linear connection, any vector field can be parallel-transported
along $\gamma$.

\begin{itemize}

\item The procedure is then very simple: take, in the way previously
discussed, a point $P$ on the curve and find the corresponding
``nullifying'' tetrad $\{H_{a}\}$; then, parallel-transport it along
the curve.

\item The components $\omega^{a}{}_{b \nu}(x)$ vanish along the curve;
this means that the components $ \Gamma^{\lambda}{}_{\mu \nu}(x)$
reduce to those of the Weitzenb\"ock connection of the tetrad
$\{H_a\}$ along the curve; the problem is equivalent to finding a
tetrad field whose Weitzenb\"ock connection coincides with $\Gamma$
along the curve.

\item Applying a point-independent Lorentz transformation to the fixed
point equation (\ref{tovanish3}) yields the same equation; applied to
a solution, it gives another solution; thus, such solutions are
defined up to fixed-point Lorentz transformations; along a curve, as
it will be seen below, the Lorentz transformations will be distinct at
each point.

\item Take  (\ref{tovanish4}) in the form
\begin{equation} 
 \ H^{b}{}_{\mu}(x) \frac{d}{du} \ H_{b}{}^{\lambda}(x) = -\ {\bar
 \Gamma}^{(H) \lambda}{}_{\mu \nu}(x) U^{\nu} = -\
 \Gamma^{\lambda}{}_{\mu \nu}(x) U^{\nu}
\end{equation}
and contract it with $U^{\mu}$:
\begin{equation} 
U^{\mu} H^{b}{}_{\mu}(x) \frac{d}{du} \ H_{b}{}^{\lambda}(x) = -\
\Gamma^{\lambda}{}_{\mu \nu}(x) U^{\mu} U^{\nu}.
\end{equation}
Seen from the tetrad, the tangent field will be $U^{b} = H^{b}{}_{\mu}
U^{\mu}$, and the above formula is
\begin{equation} 
U^{b} \frac{d}{du} \ H_{b}{}^{\lambda}(x) + \Gamma^{\lambda}{}_{\mu
\nu}(x) U^{\mu} U^{\nu} = 0,
\end{equation}
or
\begin{equation}
H_{b}{}^{\lambda}(x) \ \frac{d}{du}\ U^b = \frac{d}{du} \
U^{\lambda}(x) + \Gamma^{\lambda}{}_{\mu \nu}(x) U^{\mu} U^{\nu} =
\nabla_U U^{\lambda}(x). \label{DU}
\end{equation}

\item Seen from the frame $\{H_a = H_{a}{}^{\lambda}\ \frac{\partial
}{\partial x^{\lambda}} \}$, the equation  for  a self-parallel curve will
be 
\begin{equation}
 \frac{d}{du}\ U^a = 0. \label{noaxel}
\end{equation}

\item If $U^a = \frac{d y^a}{d \tau}$, for $y^a$ some coordinates on
Minkowski space and $d \tau^2 = \eta_{ab} d y^a d y^b$, then
$\frac{d}{du}\ U^a = \frac{d\tau}{du}\frac{d U^a}{d \tau}$ with
$\frac{d\tau}{du} \ne 0$, and $\frac{d}{du} U^a$ = 0 $\Leftrightarrow
\frac{d^2 y^a}{d \tau^2}$ = 0.
 
\item If an external force is present, then $m \frac{\nabla\ }{\nabla u} \
U^{\lambda}(x)$ = $F^{\lambda}$ and $m \frac{d}{du}\ U^a = F^{a}$.

\item The point solution given above is a first-order local solution;
there can be a global unique solution only if the connection is flat. 

\end{itemize}

Summing up: given (i) given a piecewise-differentiable curve $\gamma$,
(ii) a point $P$ on $\gamma$ with a coordinate chart $(N, x^\mu)$
around it, and (iii) the components $\Gamma^{\lambda}{}_{\mu \nu}(x)$
of a linear connection in the holonomic base defined by the coordinate
system $\{x^\mu\}$, then there exists a tetrad frame $\{H_{a}\}$ at
$P$, parallel-transported along $\gamma$, in which the connection
components vanish along the curve, as long as it traverses a small
enough neighborhood $N' \subset N$.

A generic tetrad $h_{a}$ is not parallel-transported along a
Levi-Civita geodesic of velocity $U$. Its deviation from parallelism
is measured by the spin connection. Indeed, from
Eq.(\ref{eq:omegabolasdevi}),
\begin{equation} %
\frac{d}{du} \ h_{b}{}^{\lambda} + \gammabol^{\lambda}{}_{\mu \nu}\
U^{\nu}\ h_{b}{}^{\mu} = h_{a}{}^{\lambda}\ \omegabol^{a}{}_{b \nu}\
U^{\nu}.  \label{here}
\end{equation} %
 Let us write it in still another form,
\begin{equation} %
\nabla_U  h_{b}  =  \frac{d}{d u}\ h_{b} +
\gammabol (\frac{d}{d u})
\ h_{b}  = \omegabol^{a}{}_{b}
(\frac{d}{d u})  h_{a}.
    \label{eq:hderiv} 
\end{equation} %
This holds for {\em any} curve with tangent vector $U = (\frac{d}{d
u})$.  It means that the frame $\{ h_{b}\}$ can be
parallel-transported along no curve.  The spin connection forbids it,
and gives the rate of change with respect to parallel transport.

%%%%%%%%%%%%%%%%%%%%%%%%%%%%%%%%
\subsection{Free--falling frame}
%%%%%%%%%%%%%%%%%%%%%%%%%%%%%%%%

The Principle says that it is possible to choose a frame in which the
connection vanishes.  Let us see now how to obtain such a free falling
frame $\{ H_{a}\}$ from an arbitrary initial frame.  Contracting
(\ref{eq:omegaasdevi}) with $H^{a}{}_{\lambda}$,
\begin{equation} 
H^{a}{}_{\lambda} \frac{d}{du} \ h_{b}{}^{\lambda} +
\Gamma^{\lambda}{}_{\mu \nu}\ U^{\nu}\ H^{a}{}_{\lambda} h_{b}{}^{\mu}
= H^{a}{}_{\lambda} h_{c}{}^{\lambda}\ \omega^{c}{}_{b \nu}\ U^{\nu},
\end{equation}
which is the same as
\begin{equation} 
 \frac{d}{du} \ \left(H^{a}{}_{\lambda} h_{b}{}^{\lambda}\right) -
 h_{b}{}^{\mu} \left( \frac{d}{du} \ H^{a}{}_{\mu} -
 \Gamma^{\lambda}{}_{\mu \nu}\ U^{\nu}\ H^{a}{}_{\lambda} \right) =
 H^{a}{}_{\lambda} h_{c}{}^{\lambda}\ \omega^{c}{}_{b \nu}\ U^{\nu}.
\end{equation}
The second term in the left-hand side vanishes by
Eq.(\ref{tovanish5}), so that we remain with
\begin{equation} %
\frac{d}{du} \ \left(H^{a}{}_{\lambda} h_{b}{}^{\lambda}\right) -
(H^{a}{}_{\lambda} h_{c}{}^{\lambda})\ \omega^{c}{}_{b \nu}\ U^{\nu}\
= 0.
\label{intheway}
\end{equation} %
What appears in both parentheses is a point-dependent Lorentz transformation
(\ref{goodLor}) relating the tetrad $h_{a}$ to the frame $H_{a}$ in which
$\Gamma$ vanishes:
\begin{equation} %
H^{a}{}_{\lambda}  = \Lambda^{a}{}_{b}\ h^{b}{}_{\lambda}\ , \ \ 
H^{a}   = \Lambda^{a}{}_{b}\ h^{b}. 
\end{equation} %
This means
\begin{equation} %
\Lambda^{a}{}_{b} = H^{a}{}_{\lambda}\ h_{b}{}^{\lambda} 
\label{goodLor2}
\end{equation} %
and holds on the common definition domain of the tetrad
fields. We have, of course,
\begin{equation}
\nabla_U h_a{}^\lambda = H_b{}^\lambda \nabla_U \Lambda^b{}_a.
\end{equation}
For the particular case of the Levi-Civita connection $\gammabol$,
taking (\ref{goodLor2}) into (\ref{intheway}), we arrive at a
relationship which holds on the intersection of that domain with a
geodesics:
\begin{equation} %
	\frac{d}{du} \  \Lambda^{a}{}_{b}  -
 \Lambda^{a}{}_{c}\ \omegabol^{c}{}_{b d}\ U^{d}  = 0.
	\label{eq:movingLor}
\end{equation} %
This equation gives the change, along the metric geodesic, of the
Lorentz transformation taking a tetrad $h_{a}$ into the frame $H_{a}$. 
The vector formed by each row of the Lorentz matrix is
parallel-transported along the line.  Contracting with the inverse
Lorentz transformation $(\Lambda^{-1})^{a}{}_{b} = h^a{}_{\rho}
H_b{}^{\rho}$ the expression above gives
\begin{equation} %
\omegabol^{a}{}_{b d}\ U^{d} =
(\Lambda^{-1})^{a}{}_{c}\ \frac{d}{du} \
\Lambda{}^{c}{}_{b}\ =
(\Lambda^{-1} \frac{d}{du} \
\Lambda)^{a}{}_{b}. \label{Lorvac}
\end{equation} %
This is, in the language of differential forms, 
\begin{equation} %
\omegabol^{a}{}_{b d}\ h^{d} \left( \frac{d}{du} \right) =
(\Lambda^{-1} d \Lambda)^{a}{}_{b} \left( \frac{d}{du} \right).
\end{equation} %
Thus {\em on the points of the curve}, the connection has the form of
a gauge Lorentz vacuum:
\begin{equation}
\omegabol^{a}{}_{b }  =
(\Lambda^{-1} d \Lambda)^{a}{}_{b}.
\end{equation}
This expression deserves further attention.  A Lorentz transformation,
as a member of the Lorentz group, has the form
\begin{equation} %
\Lambda = e^{W} = e^{\onehalf J_{a b}\ \alpha^{ab}}.
\end{equation}
Matrix $W$ belongs to the group Lie algebra, $J_{a b}$ are
the generators and $\alpha^{ab}$ are parameters specifying the
transformation ($\alpha^{i j}$ for rotation angles, $\alpha^{a 0}$ for
boosts parameters).  Clearly
\begin{equation}
\omegabol^{a}{}_{b }  = (\Lambda^{-1} d \Lambda)^{a}{}_{b}  =
(d \ln \Lambda)^{a}{}_{b} = (d W)^{a}{}_{b} = d W^{a}{}_{b}.
\end{equation}

The generators can be obtained in a standard way.  Introduce first the
canonical basis $\{\Delta_a{}^b\}$ for the generators of the linear
group $GL(4,\mathbb{R})$.  These matrices have entries
$(\Delta_a{}^b)_c{}^d$ = $\delta^b_c\ \delta^d_a$.  Define new
matrices with indices lowered by the Lorentz metric, $\{\Delta_{ab}\}$
= $\eta_{bc}\Delta_a{}^c$.  Then, the set of matrices $\{J_{a b} =
\Delta_{ab} - \Delta_{ba}\}$ provide a representation for the
generators of the Lorentz group.  Their entries are $(J_{cd})^a{}_b$ =
$\eta_{db}\ \delta^a_c$ - $\eta_{cb}\ \delta^a_d$.  Consequently,
\begin{equation} 
 W^{a}{}_{b} = \onehalf (J_{c d})^a{}_b \ \alpha^{c d} = 
\onehalf (\eta_{db}\ \delta^a_c - \eta_{cb}\ \delta^a_d) \ \alpha^{c d}
=  \onehalf (\alpha^{a}{}_{b} -\alpha_{b}{}^{a}) = \alpha^{a}{}_{b}.
\end{equation}
Therefore, on the points of the curve,
\begin{equation}
\omegabol^{a}{}_{b \nu}  = \partial_\nu \alpha^{a}{}_{b}.
\end{equation}
The spin connection, seen from the tetrad $h$, is given by the
derivatives of the parameters of the Lorentz transformation taking $h$
into the locally inertial tetrad $H$. Once we have learned about the
relationship of the spin connection to Lorentz transformations, we can go
back to
\begin{equation} 
\nabla_U h_{b}{}^{\lambda} = \frac{\nabla\ }{\nabla u}\
h_{b}{}^{\lambda} = \frac{d}{du}\ h_{b}{}^{\lambda} +
\gammabol^{\lambda}{}_{\mu \nu} U^{\nu} h_{b}{}^{\mu} =
h_{a}{}^{\lambda}\ \omegabol^{a}{}_{b \nu} U^{\nu} =
h_{a}{}^{\lambda}\ \frac{d}{du}\ \alpha^{a}{}_{b}.
\end{equation}
If we choose a frame such that $h_{0}{}^{\lambda} = U^{\lambda}$, the
acceleration will be
$$%
\abol^{\lambda} = \frac{d}{du}\ U^{\lambda} +
\gammabol^{\lambda}{}_{\mu \nu} U^{\mu} U^{\nu} = h_{a}{}^{\lambda}\
\omegabol^{a}{}_{0 \nu} U^{\nu} = h_{0}{}^{\lambda}\
\omegabol^{0}{}_{0 \nu} U^{\nu} + h_{j}{}^{\lambda}\
\omegabol^{j}{}_{0 \nu} U^{\nu}
$$%
\begin{equation}
=  
h_{j}{}^{\lambda}\ \omegabol^{j}{}_{0 \nu} U^{\nu}  = 
h_{k}{}^{\lambda}\ \frac{d}{du}\ \alpha^{k}{}_{0}.
\end{equation}
Only boosts turn up. This means that, on a geodesic, the
spin connection in some tetrad field $\{h_a\}$ is determined by the
special Lorentz transformation (\ref{goodLor}) taking $\{h_a\}$ into
that tetrad field $\{H_a\}$ in which it vanishes.  The spin connection
is a vacuum along each geodesic.  Boost parameters (essentially frame
velocities) and rotation angles will continuously change along the
curve to transform $h_a$ into $H_a$ and keep the connection, as seen
from $\{H_a\}$, equal to zero.  At each point of a metric-geodesic
observer, the velocity $U$ differs by a Lorentz transformation from
the velocity $U$ satisfying the forceless equation (\ref{noaxel}).

Using (\ref{Lorvac}), the geodesic equation (\ref{eq:geodesic2}) seen
from the frame $h_{a}$ takes on the form
\begin{equation}
\frac{d\ }{du}\ U^{a} + (\Lambda^{-1}\ \frac{d\ }{du}
\ \Lambda)^{a}{}_{b}\ U^{b} = 0 
\end{equation}
which, once  multiplied on the left by $\Lambda$, gives
 Eq.(\ref{noaxel}) for the present case,
\begin{equation}
\Lambda^{c}{}_{a}\ \frac{d\ }{du}\ U^{a} + \frac{d\ }{du}\
(\Lambda^{c}{}_{a})\ U^{a} = \frac{d\ }{du}\ (\Lambda^{c}{}_{a} \
U^{a}) = \frac{d\ }{du}\ {\bar U}^{c} = 0.
\end{equation}

%%%%%%%%%%%%%%%%%%%%%%%%%%%
\subsection{Real observers}
%%%%%%%%%%%%%%%%%%%%%%%%%%%

It is important to stress that the above results hold only along a
curve --- a one-dimensional domain --- so that curvature is not
probed.  Curvature, the real gravitational field strength, only
manifests itself on two-dimensional domains.  If curvature is
nonvanishing, no vector field can be parallel-transported along two
distinct lines.  In effect, consider two curves with tangent fields
$U$ and $V$, with parameters $u$ and $v$, so that
\begin{equation}
\nabla_U H^{\lambda} = U^\nu \nabla_\nu H^{\lambda} = U^\nu
[\partial_\nu H^{\lambda} + \Gamma^{\lambda}{}_{\rho \nu} U^\nu
H^\rho];
\end{equation} 
\begin{equation}
\nabla_V H^{\lambda} = V^\mu \nabla_\mu H^{\lambda} = V^\mu
[\partial_\mu H^{\lambda} + \Gamma^{\lambda}{}_{\rho \mu} V^\mu
H^\rho].
\end{equation}
Then, one finds
\begin{equation}
\left[\nabla_U \nabla_V - \nabla_V \nabla_U \right] H^{\lambda} =
R^{\lambda}{}_{\rho \mu \nu}\ H^\rho U^\mu V^\nu.
\end{equation}
This shows that, if $ \nabla_U H^{\lambda} = 0$, then forcibly
$\nabla_U \nabla_V H^{\lambda} $ = $ R^{\lambda}{}_{\rho \mu \nu}\
H^\rho U^\mu V^\nu \ne 0$.  The apparent ``turning-off'' of
gravitation is an effect of the one-dimensional character of the curve
representing the ideal observer.  A real observer will have spatial
extension and will be, consequently, represented by a bunch of
neighboring timelike curves.  Such curves will deviate from each other
in a way which depends of the curvature (and torsion, if present). 
The deviation $X^{\lambda}$ will, in effect, respect Jacobi's equation
for a general curve.  The Jacobi equation for a general connection in
a holonomic frame$^($\cite{AP95b}$^)$ is
\begin{equation} %
\nabla_U\nabla_U X^{\lambda} + R^{\lambda}{}_{\mu \rho
\nu}\ U^{\mu} X^{\rho} U^{\nu} + U^{\nu} \nabla_U
\left(T^{\lambda}{}_{\mu \nu} X^{\mu}\right) = 0.
\end{equation} %
When the curve is self-parallel ($\frac{\nabla\ }{\nabla u}\ U^{\nu} =
0$), it gives the geodesic deviation:
\begin{equation} %
\nabla_U \nabla_U X^{\lambda} + R^{\lambda}{}_{\mu
\rho \nu}\ U^{\mu} X^{\rho} U^{\nu} + \nabla_U
\left(T^{\lambda}{}_{\mu \nu} X^{\mu} U^{\nu} \right) = 0.
\end{equation} %
For a metric geodesic in the tetrad frame, the equation for the
deviation $X^{a}$ will be
\begin{equation} %
\nabla_U\nabla_U X^{a} + \Rbol^{a}{}_{b c d}\
U^{b} X^{c} U^{d} = 0.
\end{equation} %
The Equivalence Principle is not stated for real observers. These are
extended objects in the 3-dimensional space sections and, by observing
nearby curves, will always be able to detect the curvature.  Actually,
curvature will be detectable on domains of $2$ or more dimensions. 
Let us profit to make clear what should be meant by the word
``local'', by which some authors understand ``at a point in
spacetime'', others ``in a neighborhood of a point in spacetime``,
still others ``on a piece of trajectory through a point in
spacetime''.  This issue has been definitively clarified by
Iliev:$^($\cite{Ilioc}$^)$ in what concerns the Principle, only the
last meaning applies.

We have seen that:
\begin{enumerate}
\item given {\em any} connection $\Gamma$ and {\em any} differentiable
curve $\gamma$, there exists a tetrad field $H$ which is
parallel-transported by $\Gamma$ along $\gamma$;%

\item along $\gamma$, any other tetrad field is taken into $H$ by a
point-dependent Lorentz transformation;%

\item if $\gamma$ is a geodesic of $\Gamma$, then $H$ is in free fall
along $\gamma$; seen from $H$, $\gamma$ is a straight
line;

\item if $\gamma$ is a timelike geodesic of $\Gamma$, then $H$ can be
assimilated to a local inertial frame, or to an inertial ideal
observer, which will see the world as described by
Special Relativity;

\item this holds only on the points of the 1-dimensional domain
$\gamma$; a real observer, composed of bunch of curves around
$\gamma$, will sense the gravitational field.

\end{enumerate}

%%%%%%%%%%%%%%%%%%%%%%%%%%%%%%%%%%%%%%%
\section{The case of a gauge connection}
%%%%%%%%%%%%%%%%%%%%%%%%%%%%%%%%%%%%%%%
\label{vanishGaugecon}

Reduced to the statement that the connection can be made to vanish at
a point, the Equivalence Principle is not specific of gravitation.  In
effect, it has been clearly shown by Iliev$^($\cite{Ilioc4}$^)$ that
also in a gauge theory it is possible to choose a gauge to make the
connection components to vanish at a point and along a curve.  As all
the main points are already present in the abelian case, we shall
present in some detail the case of electrodynamics and only indicate
the generalization to the non--abelian case.

%%%%%%%%%%%%%%%%%%%%%%%%%%%%
\subsection{Electrodynamics}
%%%%%%%%%%%%%%%%%%%%%%%%%%%%

Consider a potential $A_{\mu}(x)$.  Take a point $P$ and a coordinate
neighborhood around $P$.  For the sake of simplicity, choose
coordinates $x^{\lambda}$ with origin at $P$, that is, such that
$x^{\lambda}(P)$ = $0$.  Suppose further that the field $F_{\mu \nu}$
is well--defined, that is, the derivatives of $A_{\mu}(x)$ are finite
in some neighborhood $N$ around $P$.  We can take $N$ as a member of
the implicit differentiable atlas.  Call $\gamma_{\mu}$ the value of
$A_{\mu}(x)$ at the point $P$, $\gamma_{\mu}$ = $A_{\mu}(P)$.  There
will be a domain $N' \subset N$ around $P$, small enough for
$A_{\mu}(x)$ to be approximated as
\begin{equation}%
A_{\mu}(x) = \gamma_{\mu} + x^{\lambda} (\partial_{\lambda}
A_{\mu})(P) = \gamma_{\mu} + x^{\lambda} \gamma'_{\mu \lambda}.
\label{AED}
\end{equation}%
What we shall do is to choose, in the general expression for a gauge
transformation
\begin{equation}
A'_{\mu}(x) = A_{\mu}(x) + \partial_{\mu} \phi(x),
\end{equation}
a very particular $\phi(x)$.  As the two constants $\gamma_{\mu}$ and
$\gamma'_{\mu \lambda}$ are given, we take on $N'$ the function
\begin{equation}
\phi(x) = - \gamma_{\rho} x^{\rho} -  \gamma'_{\rho \sigma}
x^{\rho} x^{\sigma}.
\end{equation}
Then, $\partial_{\mu} \phi(x)$ = $- \gamma_{\mu}$ - $\gamma'_{\mu
\lambda} x^{\lambda}$ - $\gamma'_{\lambda \mu} x^{\lambda}$ and,
always on $N'$,
\begin{equation} %
A'_{\mu}(x) = A_{\mu}(x) + \partial_{\mu} \phi(x) = - \ x^{\lambda}
\gamma'_{\lambda \mu}, \label{atP}
\end{equation} %
\noindent which gives $A'_{\mu}(P)$ = $0$.  As $\partial_{\mu}
A'_{\nu}(x)$ = - $\gamma'_{\mu \nu}$, $F_{\mu \nu}(P)$ = $\gamma'_{\nu
\mu} - \gamma'_{\mu \nu}$ as it should be.  Thus, given any
electromagnetic potential and a point $P$, it is always possible to
choose a gauge in which the potential vanishes at $P$.  It is possible
to go a bit beyond the above approximations, by taking for $A_{\mu}(x)
$ a Taylor expansion around $P$:
\begin{equation}
A_{\mu}(x) = A_{\mu}(P) + \sum_{j=1}^{\infty} \frac{x^{\lambda_{1}}
x^{\lambda_{2}} \ldots x^{\lambda_{j}}}{j!} \left[
\partial_{\lambda_{1}} \partial_{\lambda_{2}} \ldots
\partial_{\lambda_{j}} A_{\mu} \right]_P.
\end{equation} 
Choose then 
\begin{equation}
\phi (x) = - x^{\rho} A_{\rho} (P) - x^{\rho} 
\sum_{j=1}^{\infty} \frac{x^{\lambda_{1}}
x^{\lambda_{2}} \ldots x^{\lambda_{j}}}{(j+1)!} \left[
\partial_{\lambda_{1}} \partial_{\lambda_{2}} \ldots
\partial_{\lambda_{j}} A_{\rho} \right]_P,
\end{equation} 
so that
\begin{equation} 
\partial_{\mu} \phi (x) = -\  A_{\mu} (P)
 - \sum_{j=1}^{\infty} \frac{x^{\lambda_{1}}
x^{\lambda_{2}} \ldots x^{\lambda_{j}}}{(j+1)!} \left[
\partial_{\lambda_{1}} \partial_{\lambda_{2}} \ldots
\partial_{\lambda_{j}} A_{\mu} \right]_P  -  
\sum_{j=1}^{\infty} j\ \frac{x^{\lambda_{1}}
x^{\lambda_{2}} \ldots x^{\lambda_{j-1}}}{(j+1)!} \left[
\partial_{\lambda_{1}} \partial_{\lambda_{2}} \ldots
\partial_{\lambda_{j-1}} \partial_{\mu}A_{\rho} \right]_P.
\end{equation} 
We have then
\begin{equation}
A'_{\mu}(x) = A_{\mu}(x) + \partial_{\mu} \phi(x) =\onehalf x^{\rho}
F_{\rho \mu}(P) + \sum_{j=2}^{\infty} \frac{j}{(j+1)!} x^{\lambda_{1}}
x^{\lambda_{2}} \ldots x^{\lambda_{j}}\left[ \partial_{\lambda_{1}}
\partial_{\lambda_{2}} \ldots \partial_{\lambda_{j-1}} F_{\lambda_{j}
\mu} \right]_P.
\end{equation}
From which again $A'_{\mu}(P)$ = $0$, but further $[\partial_{\rho }
A'_{\mu} - \partial_{\mu} A'_{\rho}]_P$ = $F_{\rho \mu}(P)$.

%%%%%%%%%%%%%%%%%%%%%%%%%%%%%%
\subsection{Non--abelian case}
%%%%%%%%%%%%%%%%%%%%%%%%%%%%%%

Considerations analogous to those above will keep holding in what
concerns coordinates and neighborhoods.  The generic gauge
transformation is now
\begin{equation}%
{\bf A}'_{\mu}(x) = {\bf g}^{-1}(x) {\bf A}_{\mu}(x) {\bf g}(x) + {\bf
g}^{-1}(x) \partial_{\mu} {\bf g}(x) = {\bf g}^{-1}(x) {\bf
A}_{\mu}(x) {\bf g}(x) + \partial_{\mu} \ln {\bf g}(x).
\end{equation}%
The main difference comes from the fact that now matrices (${\bf A}$,
${\bf g}$, ${\boldsymbol \gamma}$, ${\bf F}$) are at work.  
Choose in some $N'$ the same form of (\ref{AED}), ${\bf
A}_{\mu}(x)$ = ${\boldsymbol \gamma}_{\mu}$ + $x^{\lambda}
{\boldsymbol \gamma}'_{\mu \lambda}$.  Take
\begin{equation}%
\ln {\bf g}(x) = -\ {\boldsymbol \gamma}_{\rho} x^{\rho} - 
{\boldsymbol \gamma}^{\prime}_{\rho \sigma}\ x^{\rho} x^{\sigma}.
\end{equation}%
It follows that 
\begin{equation}
{\bf g}(x) \approx \ I - \ {\boldsymbol \gamma}_{\rho} x^{\rho} -
{\boldsymbol \gamma}^{\prime}_{\rho \sigma}\ x^{\rho} x^{\sigma} +
\onehalf \ {\boldsymbol \gamma}_{\rho} x^{\rho} {\boldsymbol
\gamma}_{\sigma} x^{\sigma} \ ; \ {\bf g}^{-1}(x) \approx \ I + \
{\boldsymbol \gamma}_{\rho} x^{\rho} + {\boldsymbol
\gamma}^{\prime}_{\rho \sigma}\ x^{\rho} x^{\sigma} \ + \onehalf \
{\boldsymbol \gamma}_{\rho} x^{\rho} {\boldsymbol \gamma}_{\sigma}
x^{\sigma},
\end{equation}
\begin{equation}%
{\bf A}'_{\mu}(x) = -\ {\boldsymbol \gamma}'_{\lambda \mu} x^{\lambda}
+[{\boldsymbol \gamma}_{\lambda}, {\boldsymbol \gamma}_{\mu}]\
x^{\lambda}.
\end{equation}%
This gives back (\ref{atP}) in the abelian case.  We see that the
gauge transformed ${\bf A}'_{\mu}(P)$ = $0$, so that the original
${\bf A}_{\mu}(P)$ ``touches'' a vacuum at the point $P$, ${\bf
A}_{\mu}(P)$ = ${\bf g}\ \partial_{\mu} {\bf g}^{-1}$.  The covariant
field strength remains what it should be:
\begin{equation}
{\bf F}_{\mu \nu}(P) = {\boldsymbol \gamma}'_{\nu \mu } - {\boldsymbol
\gamma}'_{\mu \nu} + {\boldsymbol \gamma}_{\mu} {\boldsymbol
\gamma}_{\nu} - {\boldsymbol \gamma}_{\nu} {\boldsymbol \gamma}_{\mu}.
\end{equation}
We insist that the vanishing of ${\bf A}'_{\mu}$ takes place at one
point. If ${\bf A}_{\mu}(x)$ = ${\bf g}\ \partial_{\mu} {\bf g}^{-1}$
in an open domain, however small, then ${\bf F}_{\mu \nu}(x) = 0$ in
the domain.  

To obtain the vanishing of the gauge potential along a curve, it is
necessary to proceed by infinitesimal steps. The procedure would be
as follows:
\begin{itemize}
\item start at a point P on the curve $\gamma$ with $A^{a}{}_\mu =
0$;

\item take a neighboring point Q on $\gamma$, separated from P by a
variation $du$ in the parameter; at Q,
\[
A^{\prime a}{}_\mu
\frac{dx^\mu}{du} du = A^{a}{}_\mu \frac{dx^\mu}{du} du - \nabla_\mu
W^{a} \frac{dx^\mu}{du} du = 0 - \frac{\nabla\ }{du} W^{a} du,
\] or $A^{\prime a}{}_\mu = -\ \nabla_\mu W^{a}$, a
pure gauge;

\item start now  from Q, and repeat the procedure, taking another point $Q'$
on $\gamma$; and so on.
\end{itemize}
This amounts to a step-by-step definition of an ordered product and
says that, along any differentiable curve traversing a domain $N'$,
there is a continuous choice of gauges in which the potential
vanishes.  Notice that this only holds for a theory which keeps on
gauge invariance.  Except for the abelian electromagnetic sector,
gauge symmetry is broken in electroweak theory.  A gauge is chosen
once and for all by the extra scalar field, so as to provide the boson
masses.  No choice of gauge is left, and the boson fields cannot be
annulated.

%%%%%%%%%%%%%%%%%%%%%%%%%%%%%
\section{The force equation }
%%%%%%%%%%%%%%%%%%%%%%%%%%%%%
\label{sec:force}

A massive particle without additional structure (for instance,
supposing that the effect of its spin is negligible, or zero) will
follow the geodesic equation.  A charged test particle
in a gravitational and an electromagnetic field will respect
instead the Lorentz law of force in the presence of a gravitational field
\begin{equation}
mc\ \nabla_U U^{\lambda} =  m c
\left(\frac{d}{du}\ U^\lambda + \Gamma^\lambda{}_{
\mu \rho} U^\mu U^\rho\right)
= \frac{e}{c}\ F^\lambda{}_{\mu}  U^\mu.
\end{equation}

Gravitation, represented by the term in $\Gamma$, is an inertial
force, because that term can be made to vanish at each point by a
choice of reference system.  The same holds if we think of a test
field, for example a vector field $\phi^{\alpha}$.  It couples to
gravitation through the $\Gamma$ term in the minimal coupling
prescription $\partial_{\gamma} \phi^{\alpha} \Rightarrow
\partial_{\gamma} \phi^{\alpha}$ + $\Gamma^{\alpha}{}_{\beta \gamma}
\phi^{\beta}$. It decouples at a point, by the same choice of
reference system. This can, as seen above, be generalized to a curve.

To fix the ideas, let us consider the paradigm of a gauge theory, the
original Yang--Mills $SU(2)$ model.  A test field $\phi^{a}$ carrying
isospin will belong to some representation of $SU(2)$ with generators
$T_{a}$, and will feel the presence of a Yang--Mills field through the
minimal coupling prescription $\partial_{\gamma} \phi^{a} \Rightarrow
\partial_{\gamma} \phi^{a}$ + $A^{b}{}_{\gamma} T_{b} \phi^{a}$. 
Matrices $g = \exp [\omega^{a} T_{a}]$, representing $SU(2)$
transformations will act on the (internal) vector space (or module)
$V$ to which $\phi^{a}$ belongs.  A gauge transformation can be
conceived in the usual, active way, as a change in $\phi^{a}$.  But it
can also be conceived in the passive way, as a change of frame in $V$. 
In that case, a choice of gauge corresponds to a choice of frame.  And
we have seen that, fixed a point $P$ and a gauge potential ${\bf
A}_{\mu}$, it is possible to choose a frame in $V$ such that
$\phi^{a}$ decouples from ${\bf A}_{\mu}$ at $P$. It is doubtful that
we can use the word ``inertia'' in this internal case, but the
possibility of zeroing the connection is a common feature of all gauge
interactions.  As already remarked, this is not true for the
broken-symmetry electroweak interactions.

The motion of a test particle of mass $m$ submitted to an $SU(2)$
Yang--Mills field is described by (i) its spacetime coordinates and
(ii) an ``internal'' vector ${\bf I} = \{I_{a} \}$ giving its state in
isotopic space.  The corresponding dynamic equations$^($\cite{Won70}$^)$
are (i) the generalized Lorentz force law
\begin{equation}%
mc\ \nabla_U U^{\lambda} = \frac{1}{c}\ I_{a} \; F^{a \lambda}{}_{\nu}
\; U^{\nu} = \frac{1}{c}\ I_{a} \; (\partial^{\lambda} A^{a}{}_{\nu} -
\partial_{\nu} A^{a}{}^{\lambda} + c^a{}_{bc}
A^{b}{}^{\lambda}A^{c}{}_{\nu}) \; U^{\nu},
\label{Lorforce2}
\end{equation} %
and (ii) the  charge--precession, or Wong equation,
\begin{equation}%
\nabla_U \; {\bf I} \; + {\bf A}_{\mu} \; 
U^{\mu}\times {\bf I}  \; = \; 0. 
\end{equation} %
This equation says that internal motion is a parallel--transport by
the internal connection and, furthermore, a precession (because ${\bf
I}^{2}$ is conserved).  For unbroken models it is possible to choose,
at each point, a gauge in which there is no charge--precession.  The
force, however, does not change.  The vanishing of $A$ along a curve
would mean a gauge $A_\mu U^\mu (s) = 0$, which would generalize to
general spaces the well-known gauge $A_\mu n^\mu = 0$ used in
Minkowski space.  That would, however, lead only to absence of
precession, not of force (though only the abelian, derivative part of
$F$ would contribute).  Inertial and non-inertial forces are clearly
distinguished.  Gauges are internal contrivances, while frames
participate in the very structure of the habitat of each physical
object and the scene of physical process, spacetime.  Furthermore,
Physics is {\em not} invariant under changes of frames, unless they
are related by Lorentz transformations and consequently induce the
same metric.  An ideal observer in a gravitational field is locally
equivalent to an ideal observer in the absence of gravitation, while
an ideal observer in a gauge field will always feel its presence.  At
least two ideal observers are needed to detect gravitation, but only
one is enough to detect an electromagnetic field.  In this sense gauge
fields are local, and gravitation is not.

Concerning the Quantum Mechanics of a system immersed in a
gravitational background, an ideal observer --- a point in 3-space ---
is indeterminate.  Quantum Mechanics in 3-space will always probe a
3-dimensional domain, intersecting a bunch of curves in spacetime and,
consequently, will always be aware of a gravitational field, however
small its effect may be.

\begin{acknowledgments}
The authors would like to thank FAPESP-Brazil and CNPq-Brazil for
financial support.
\end{acknowledgments}

\end{document}